\documentclass[aps,twocolumn,pra,superscriptaddress,amsmath,showpacs,tightenlines,pdflatex,longbibliography]{revtex4-1}
\usepackage{amssymb}
\usepackage{amsmath}
\usepackage{dcolumn}
\usepackage{graphicx}
\usepackage{mathrsfs}
\usepackage{appendix}
\usepackage{graphicx}
\usepackage{booktabs}
\usepackage{color}

\def \hide#1{}

\usepackage{url}
\usepackage[colorlinks]{hyperref}
\hypersetup{%
    plainpages=true,
    breaklinks=true,
    hypertexnames=false,
    pageanchor=true,
    colorlinks=true,
    linkcolor={blue},
    citecolor={red},
    urlcolor={blue},
    anchorcolor={black}
}
\begin{document}
\title{Ideal Quantum Nondemolition Readout of a Flux Qubit Without Purcell Limitations}

\author{Xin Wang}
\affiliation{Institute of Quantum Optics and Quantum Information,
School of Science, Xi'an Jiaotong University, Xi'an 710049, China}
\affiliation{Theoretical Quantum Physics Laboratory, RIKEN Cluster
for Pioneering Research, Wako-shi, Saitama 351-0198, Japan}

\author{Adam Miranowicz}
\affiliation{Theoretical Quantum Physics Laboratory, RIKEN Cluster
for Pioneering Research, Wako-shi, Saitama 351-0198, Japan}
\affiliation{Faculty of Physics, Adam Mickiewicz University,
61-614 Pozna\'n, Poland}

\author{Franco Nori}
\affiliation{Theoretical Quantum Physics Laboratory, RIKEN Cluster
for Pioneering Research, Wako-shi, Saitama 351-0198, Japan}
\affiliation{Physics Department, The University of Michigan, Ann
Arbor, Michigan 48109-1040, USA}
\date{\today}

\begin{abstract}
Dispersive coupling based on the Rabi model with large detuning is
widely used for quantum nondemolition (QND) qubit readout in
quantum computation. However, the measurement speed and fidelity
are usually significantly limited by the Purcell effects, i.e.:
Purcell decay, critical photon numbers, and qubit-dependent Kerr
nonlinearity. To avoid these effects, we propose how to realize an
ideal QND readout of a gradiometric flux qubit with a tunable gap
via its non-perturbative dispersive coupling (NPDC)
to a frequency-tunable measurement resonator. We show that this
NPDC-based readout mechanism is free of dipole-field
interactions, and that the qubit-QND measurement is not
deteriorated by intracavity photons. Both qubit-readout speed and
fidelity can \emph{avoid the Purcell limitations}. Moreover,
NPDC can be conveniently turned on and off via an external
control flux. We show how to extend this proposal to a multi-qubit
architecture for a joint qubit readout.
\end{abstract}

 \maketitle

\section{Introduction}
Performing large-scale quantum computation requires fast and
high-fidelity qubit-readout to compete with the decoherence of
fragile quantum
states~\cite{DiVincenzo2009,Ashhab09,Kelly2015,Terhal15}. For some
robust error-correction proposals (like surface
codes~\cite{Raussendorf07,Barends2014}), the qubit measurement
also has to be repeated many times to give an accurate diagnosis
of the corrections within a given coherence time~\cite{Fowler12}.
In quantum computing, based on circuit quantum electrodynamics
\cite{You2003,Gu2017}, a superconducting qubit is often readout by
its dispersive coupling with an auxiliary cavity~\cite{Blais04,
Mallet2009, Kockum12, Gustavsson13, Diniz2013, Wallraff05,
Bultink2018}. The cavity frequency depends on the qubit
state~\cite{Gu2017}. Applying a coherent drive to the initially
empty cavity near its resonance frequency, the qubit state is
encoded in the output field, and obtained by distinguishing two
pointer states in phase-space ~\cite{Didier15}.

Recently, a number of studies were devoted to quantum
information processing with flux
qubits, showing the increasing usefulness of these
circuits~\cite{Wu2018, Rosenberg17, Yan18, Goetz18, Orgiazzi16,
Peltonen18, Armata17}.
However, the dispersive coupling of a flux qubit is usually based
on the dipole-field interaction between a resonator and a qubit,
described by $H_{x}=g_{x}(a+a^{\dag})\sigma_{x}$~\cite{Blais04},
with $a$ ($a^{\dagger}$) being the annihilation (creation)
operator of the cavity, and $\sigma_{x,z}$ the Pauli operators of
the qubit. In the large-detuning regime, the system Hamiltonian
can be written as (hereafter we set
$\hbar=1$)~\cite{Boissonneault08,Zueco09,Boissonneault09}:
\begin{equation}
H_{\text{d}}=\omega_{r}a^{\dag}a+\frac{\omega_{q}}{2}\sigma_{z}+\chi_{z}^{I}\sigma_{z}a^{\dag}a+K_{I}(a^{\dag}a^{\dag}aa)\,\sigma_{z},
\label{induced}
\end{equation}
where $\omega_{q}$ ($\omega_{r}$) is the qubit (resonator)
frequency, and $\chi_{z}^{I}\simeq\lambda g_{x}$ is the
\emph{induced dispersive coupling} (IDC) strength with
$$\lambda=g_{x}/(\omega_{q}-\omega_{r})\ll 1.$$ The Kerr
nonlinearity $K_{I}=-\lambda^{3}g_{x}$ depends on the qubit
state. The original Hamiltonian
$H_{x}$ does not commute with the Pauli operator $\sigma_{z}$, and
therefore, the qubit readout via  $H_{\text{d}}$ is not an ideal
quantum nondemolition (QND) measurement~\cite{Didier152}.

The IDC sets limitations to both qubit measurement fidelity and
speed. First, the homodyne-detection speed relies on a high value
of the photon-escape rate $\kappa$~\cite{Houck08,Jeffrey14}.
However, due to virtual excitation exchange, this strong
dissipation channel leads to an additional qubit Purcell decay at
rate
$\Gamma_{p}=\lambda^{2}\kappa$~\cite{Houck08,Liberato14,Govia17,Gard2018},
which might destroy both gate-operation and readout fidelities
(see Appendix~A). One can suppress this additional decay by
employing a Purcell filter~\cite{Jeffrey14,Sete15,Walter17};
which, however, increases experimental complexity. Second, to
suppress the qubit error transitions induced by the dipole-field
interaction, the intracavity photon number $\langle
a^{\dag}a\rangle$ should be lower than the critical photon number
$n_{c}=1/(4\lambda^{2})$ (i.e., in the quasi-QND
regime)~\cite{Blais04}, which can lead to a poor pointer-state
separation with a long measurement time.

All these trade-off relations result from the dipole-field
interaction, which reduces the fidelity of a QND measurement.
We present here a method for measuring a gradiometric flux
qubit via its non-perturbative dispersive coupling (NPDC) with a
frequency-tunable resonator.  This mechanism results from the
\emph{dispersive coupling via a longitudinal degree of freedom of
the qubit}, rather than from perturbation theory in the
dipole-field interaction. Therefore, The intracavity photons
cannot deteriorate the QND qubit readout, and the Purcell effects
are effectively avoided. We prove that both readout fidelity and
speed can go beyond the Purcell limitations. Note that the recent
preprint~\cite{Gard2018} has also discussed the problem how to
realize a QND measurement by using transverse couplings. However,
to avoid Purcell effects, the method of Ref.~\cite{Gard2018}
requires numerical optimization of various system parameters,
including designing time-dependent coupling pulses of ultra-short
durations. The readout channel of that method also needs to be
rapidly switched on and off. All these requirements might be
challenging in experiments. In contrast to that method, our
proposal is based on the NPDC readout mechanism and, thus, it does
not suffer from such limitations.

\section{Non-perturbative dispersive coupling}

\begin{figure}[tbp]
\centering
\includegraphics[width=8.8cm]{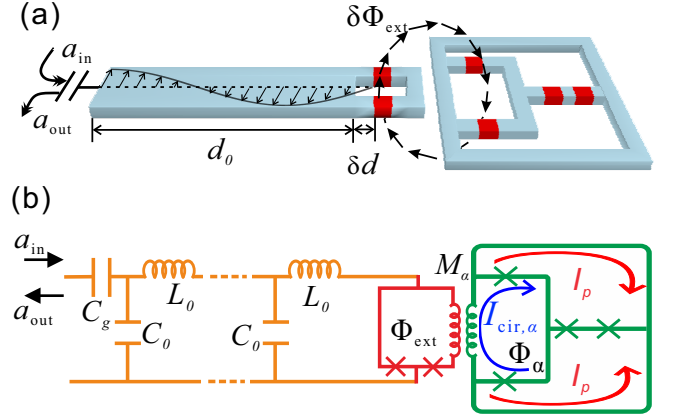}
\caption{(a) Schematic and (b) lumped circuit diagrams of a
SQUID-terminated $\lambda/4$ resonator interacting with a
gradiometric four-Josephson-junction flux qubit~\cite{Paauw09} via
the mutual inductance $M_{\alpha}$. This interaction is due to
NPDC as follows: $I_{\text{cir},\alpha}$ (related to the qubit
operator $\sigma_{z}$) generates a flux perturbation on the bias
flux $\delta \Phi_{\text{ext}}$, which in turn changes the
effective length $d_{0}$ of the resonator by an amount $\delta d$.
The result is a dispersive interaction where the photon-number
operator $a^{\dagger}a$ couples to the qubit-state operator
$\sigma_{z}$. The resonator capacitance and inductance per unit
length are $C_{0}$ and $L_{0}$, respectively. The red bars in (a)
and crosses in (b) represent the Josephson junctions. To
dispersively readout the qubit state, one can apply an input field
$a_{\text{in}}$ via the capacitance $C_{g}$. The output field is
denoted by $a_{\text{out}}$.} \label{fig1m}
\end{figure}

As demonstrated in Fig.~\ref{fig1m}, we consider a gradiometric
four-Josephson-junction (JJ) flux qubit with a tunable
gap~\cite{Fedorov10,You2008,Paauw09,Paauwthesis,Schwarz2013}
interacting with a frequency tunable transmission line resonator.
There are two kinds of circulating currents in the qubit: (i) the
conventional persistent current $I_{p}$ (red arrows) in the main
loop~\cite{You072,You2007,You2011,Xiang13,Mooij99,Orlando1999},
and (ii) the circulating current $I_{\text{cir},\alpha}$ (blue
arrow) in the $\alpha$-loop~\cite{Paauwthesis,Wang11}, which is
related to the longitudinal degree of freedom and much less
discussed in previous
studies~\cite{Stassi18,Schwarz15doc,Lambert18}. At the optimal
point, the qubit frequency $\omega_{q}$ is tuned by the flux
$\Phi_{\alpha}$ through the $\alpha$-loop. As discussed in
Appendix~C, in the Pauli-operator notation of the qubit ground and
excited state basis, the main loop current operator is
$I_{p}\sigma_{x}$, he $\alpha$-loop current operator is
$I_{\text{cir},\alpha}=I_{+}I_{0}+I_{-}\sigma_{z}$ , where $I_{0}$
is the identity operator. Note that $I_{-}$ is the difference of
the $\alpha$-loop circulating currents, which depend on the ground
and excited qubit states. This mechanism enables a
$\sigma_{z}-$type interaction (i.e., the longitudinal coupling).

As shown in Refs.~\cite{Johnson2011,Kakuyanagi2013}, if there is a
mutual inductance $M_{p}$ between the main loop and the
superconducting quantum interference device (SQUID) of the
resonator, one can also couple the $\sigma_{x}$ operator with the
resonator via the persistent current $I_{p}$. The corresponding
coupling is
\begin{equation}
H_{p}=\chi_{x}^{D}\sigma_{x}a^{\dag}a, \quad \chi_{x}^{D}=RM_{p}I_{p}.
\end{equation}
A similar type of interaction has been discussed in
Refs.~\cite{Nakano09, Kakuyanagi2013, Kakuyanagi2015}, where
quantum Zeno effects and qubit-projective measurements were
demonstrated with a flux qubit based on three Josephson junctions.
Note that $H_{p}$ does not commute with $\sigma_{z}$, and,
therefore, cannot be employed for QND measurements at the
degeneracy point. To readout a given qubit state, one should
adiabatically tune the main-loop flux far away from the degeneracy
point without damaging the qubit
state~\cite{Fedorov10,Schwarz15doc}. However, this method suffers
from a quick qubit dephasing (away from the degeneracy point) and
extra adiabatic operating steps. In our discussions, we focus on
the QND measurement based on $H_{\alpha}$.

As shown in Fig.~\ref{fig1m}, we employ a resonator terminated by
a SQUID~\cite{Wallquist06,Sandberg2008,Johansson09,Johansson09L,
Wilson2011,Johansson14,Eichler2014,Pogorzalek17} to detect the
quantized current $\hat{I}_{\text{cir},\alpha}$. The resonator is
open ended on its left side, while is terminated to ground via the
SQUID on its right side. The two JJs of the SQUID are symmetric
with identical Josephson energy $E_{s0}$ and capacitance $C_{s}$,
and the effective Josephson energy of the SQUID is tuned by the
external flux $\Phi_{\text{ext}}$ according to
$E_{s}=2E_{s0}\cos(\pi \Phi_{\text{ext}}/\Phi_{0})$, where
$\Phi_{0}$ is the flux quantum. The SQUID has a tunable nonlinear
inductance
$L_{s}(\Phi_{\text{ext}})=\Phi_{0}^{2}/[4\pi^{2}E_{s}(\Phi_{\text{ext}})]$~\cite{Johansson09,Johansson09L,Wilson2011,Johansson14,Eichler18}.

Both the SQUID nonlinear inductance $L_{s}(\Phi_{\text{ext}})$ and
the capacitance $C_{s}$ are much smaller than the total
capacitance $C_{t}=d_{0}C_{0}$ and inductance $L_{t}=d_{0}L_{0}$
of the resonator. Based on a distributed-element model and its
boundary conditions (see Appendix~B), the resonator fundamental
mode is of quarter-wavelength ($\lambda/4$) and its eigenfrequency
depends on the SQUID nonlinear inductance
$L_{s}(\Phi_{\text{ext}})$, and can be tuned via
$\Phi_{\text{ext}}$ according to the following relations:

\begin{subequations}
\begin{gather}
\omega_{r0}=\omega_{0}\left[1-\frac{L_{s}(\Phi_{\text{ext}}^{0})}{L_{t}}\right],  \\
R=\frac{\partial \omega_{r}}{\partial
\Phi_{\text{ext}}}\Big|_{\Phi_{\text{ext}}^{0}}
=-\frac{\pi\omega_{0}L_{s}(\Phi_{\text{ext}}^{0})}{\Phi_{0}L_{t}}\tan\left(\frac{\pi
\Phi_{\text{ext}}^{0}}{\Phi_{0}}\right),
\end{gather}
\label{Randfre}
\end{subequations}
\begin{figure*}[tbp]
    \centering
    \includegraphics[width=16.0cm]{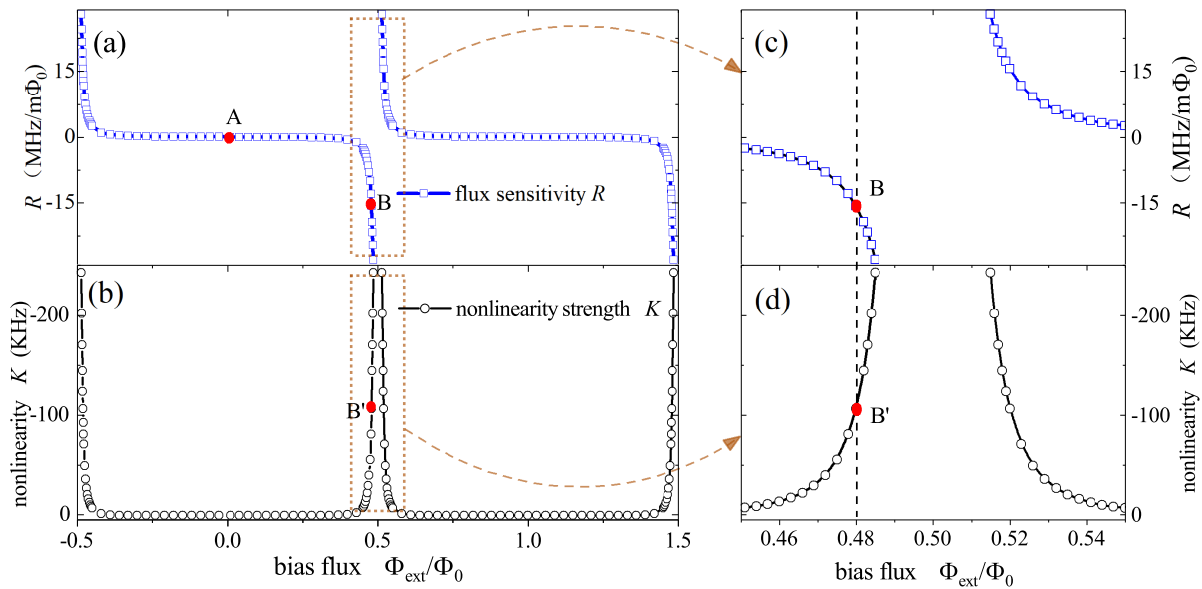}
    \caption{(a) Resonator flux sensitivity $R$ and (b) its Kerr nonlinearity
        $K$ versus the SQUID flux bias $\Phi_{\text{ext}}$. Parameters are
        adopted from the experiments in
        Ref.~\cite{Pogorzalek17,Eichler18}: $\omega_{0}/(2\pi)=
        6~\text{GHz}$, $E_{s}/(2\pi)= 2.5~\text{THz}$ and
        $L_{r}=10~\text{nH}$. Point A (B) corresponds to the
        qubit-resonator decoupling (coupling) regime.
        (c) and (d) are, respectively, the local enlarged plots of (a) and (b) in the range $\Phi_{\text{ext}} \in [0.45, 0.55]$.
    } \label{fig2m}
\end{figure*}
where we assume that the external flux $\Phi_{\text{ext}}$ is
composed of a prebiased static part $\Phi_{\text{ext}}^{0}$ and a
small deviation part $\delta \Phi_{\text{ext}}\ll
\Phi_{\text{ext}}^{0}$.  Similar to the discussions in Ref.~\cite{Johansson14}
and its experimental realization in Ref.~\cite{Eichler18},
this parametric boundary condition changes the resonator effective
length $d_{0}$ slightly, which is akin to a moving mirror for
modulating the effective wavelength in the optomechanical system.
Note that $R$ is the sensitivity of the
frequency $\omega_{r}$ tuned by the external flux
$\Phi_{\text{ext}}$, and $\omega_{r0}$ is the renormalized mode
frequency. As discussed in
Refs.~\cite{Wallquist06,Bourassa12,Eichler2014}, the attached
SQUID introduces a \emph{Kerr nonlinearity} ($K_{D}$) to the whole
circuit, which is proportional to
$[L_{s}(\Phi_{\text{ext}})/L_{t}]^{3}$, and approximately given
as~\cite{Eichler18}:
\begin{equation}
K_{D}\approx -\frac{\pi
e^{2}\omega_{r0}^{2}L_{t}}{8}\left[\frac{\pi
L_{s}(\Phi_{\text{ext}})}{2L_{t}}\right]^{3}.
\end{equation}
Figure~\ref{fig2m} shows the flux sensitivity $R$ and the Kerr
nonlinearity $K_{D}$ versus the applied flux $\Phi_{\text{ext}}$.
One finds that, when biasing $\Phi_{\text{ext}}$ from zero to
$\approx\Phi_{0}/2$, both $|R|$ and $K_{D}$ increase rapidly from
zero. This indicates that the SQUID is a highly nonlinear element
and can be exploited for \emph{enhancing nonlinear couplings}. In
our proposal, the static flux bias $\Phi_{\text{ext}}^{0}$ is
prebiased by an external field, while the flux deviation $\delta
\Phi_{\text{ext}}$ is generated by the circulating current
$I_{\text{cir},\alpha}$ of the flux qubit. As shown in Appendix~D,
one can employ the SQUID-terminated resonator to detect the qubit
state, and the Hamiltonian for this system becomes
\begin{equation}
    H_{\alpha}=\omega_{r}'a^{\dag}a+\frac{\omega_{q}}{2}\sigma_{z}+\chi_{z}^{D}\sigma_{z}a^{\dag}a+K_{D}(a^{\dag}a^{\dag}aa),
        \label{hp}
\end{equation}
where $\chi_{z}^{D}=RM_{\alpha}I_{-}$ is the NPDC strength. The
identity matrix term in $I_{\text{cir},\alpha}$ only slightly
renormalizes the mode frequency as
$\omega_{r}'=\omega_{r0}+RM_{\alpha}I_{+}$. Note that $H_{\alpha}$
commutes with $\sigma_{z}$, indicating that a qubit readout via
$H_{\alpha}$ is not deteriorated by intracavity photons.
Apparently, compared with $H_{\text{d}}$ [Eq.~(\ref{induced})]
based on the Rabi model, $H_{\alpha}$ has no relation to the
dipole-field coupling but results from the circulating current
$I_{\text{cir},\alpha}$ of the flux qubit affecting the effective
length of the resonator. The Purcell decay and critical measuring
photon-number limitation is effectively eliminated. Moreover, we
can neglect higher-energy-level transitions, because the described
NPDC is induced without dipole-field interactions. As a result,
both quantum information processing and qubit-readout fidelities can be improved compared to
other methods affected by higher-energy-level transitions. In
fact, this is the core mechanism and advantage of the QND readout
based on $H_{\alpha}$ in our proposal.

\begin{figure}[tbp]
    \centering
    \includegraphics[width=8.4cm]{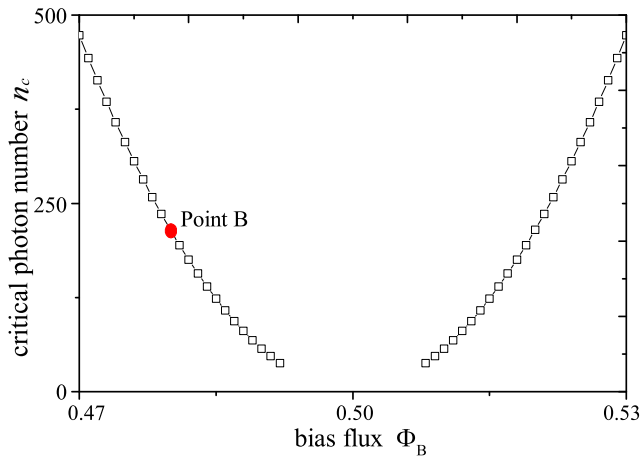}
\caption{Critical photon number $n_{c}$ versus the bias flux
$\Phi_{\text{ext}}$. The red point corresponds to the point B in
Fig.~\ref{fig2m}. The parameters assumed here are the same as
those in Fig.~\ref{fig2m}.} \label{fig3m}
\end{figure}

However, in the dispersive-readout experiments, because the
readout resonator is a nonlinear circuit element, one cannot
inject plenty of photons without any limitation. As discussed
in Appendix~B, we have assumed that the attached SQUID is
approximately a harmonic element, which corresponds to adopting
the quadratic approximation for the SQUID potential. We have
expanded its cosine potential as a quartic function. This
approximation leads to the following critical intracavity photon
number:
\begin{equation}
n_{c}=|\alpha_{c}|^{2}, \qquad  \alpha_{c}=\frac{\Phi_{0}\sqrt{{2\omega_{r}C_{t}}}}{4\pi \cos(\frac{\pi\omega_{r}}{2\omega_{0}})},
\label{Ncrt}
\end{equation}
above which the dispersive readout process cannot be realized
effectively. In Fig.~\ref{fig3m}, we plotted $n_{c}$ versus
the bias flux $\Phi_{\text{ext}}$. The red point corresponds to
that in Fig.~2. We find that the critical number decreases quickly
when $\Phi_{\text{ext}}\rightarrow 0.5$. This sets an upper bound
when employing this circuit layout for the flux qubit readout, as
will be discussed in the next section.

Another advantage of this NPDC-based layout is that the dispersive
coupling $\chi_{z}^{I}$ can be switched on/off by tuning
$\Phi_{\text{ext}}$. When implementing a gate operation to the
qubit, one can bias the flux at $\Phi_{\text{ext}}=N\Phi_{0}$
(where $N$ is an integer, see point A in Fig.~\ref{fig2m}(a), the
flux sensitivity $R$ is zero so $\chi_{z}^{D}=0$. The readout
resonator \emph{decouples} from the qubit and does not disturb
quantum information processing. Once a qubit readout is required,
one can reset $\Phi_{\text{ext}}$ around $N\Phi_{0}/2$ (point B)
to reestablish the coupling, which is fast and only takes several
nanoseconds according to \cite{Sandberg2008}.

The longitudinal degree of freedom of a flux qubit for
quantum information processing was analyzed in,
e.g.,~\cite{Richer16,Richer17,Billangeon2015b,Billangeon15}, including gate operations without
Purcell limitations. Another related work~\cite{Didier15} achieved
fast qubit readout via parametric modulation of the longitudinal
coupling. Compared with our methods, parametric modulation shifts
the qubit frequency in a time-dependent manner with a large
amplitude. Moreover, higher-order effects, which are induced by
the modulation process, might also destroy the readout fidelity of
the system in~\cite{Didier15}. Other two recent
papers~\cite{Ikonen2019,Touzard2019} on dispersive coupling are
based on a Jaynes-Cummings-type Hamiltonian. Therefore, the
Purcell effects (although could be suppressed by the Purcell
filters) still exist and limit the qubit-readout fidelity of the
systems of~\cite{Ikonen2019,Touzard2019}. Compared with these
works, \emph{our methods can avoid the Purcell effects effectively.}

For the qubit, the circulating current difference $I_{-}$ in the
$\alpha$-loop is about one order lower than $I_{p}$. At point B in
Fig.~\ref{fig2m}(c) and (d), the flux sensitivity is
$|R|/(2\pi)\simeq16~\text{MHz}/(\text{m}\Phi_{0})$ (point B) with
a nonlinearity $K_{D}/(2\pi)\simeq110~\text{kHz}$  (point
$\rm{B}'$). To avoid this Kerr nonlinear effect, one must ensure
that $\chi_{z}^{D}\gg K_{D}$.

The mutual inductance between two circuit elements can be (1)
geometric, (2) kinetic, and (3) nonlinear Josephson inductance.
Usually the geometric inductance is very small. However, the
kinetic inductance can be very large by sharing a nanowire between
two circuit elements~\cite{Zhang2019, Grnhaupt2018, Niepce2019}.
The kinetic inductance increases when decreasing the cross-section
area of a nanowire made from aluminum films. To achieve stronger
$\chi_{z}^{D}$, one can employ the kinetic mutual inductance
$M_{\alpha}$ by sharing a branch of the $\alpha$-loop with the
resonator
SQUID~\cite{Paauwthesis,Meservey1969,Annunziata2010,Natarajan2012,Doerner2018}.
As discussed in the experimental paper~\cite{Schwarz15doc}, the
kinetic inductance per unit length can be $L_{k}=5$\,pH/$\mu$m for
the cross-sectional area $S=100\times 90\,$nm$^2$. Therefore, the
value of 15pH can be easily achieved and the NPDC strength is
about $\chi_{z}^{D}/(2\pi)\backsimeq7~\text{MHz}$, which is of the
same order as the IDC strength reported in
experiments~\cite{Majer2007,Jeffrey14}, and strong enough for a
qubit QND readout.

Another method is to employ a Josephson junction as a mutual
inductance. In Ref.~\cite{Grajcar06}, the Josephson inductance was
reported as large as 40\,pH. In an experimental realization, one
can employ a much larger inductance to achieve even much stronger
NPDC than that estimated in our paper. Note that the coupling
strength $\chi_{z}^{D}$ can still be enhanced by reducing the wire
cross-section area of the kinetic inductance, or by inserting a
nonlinear JJ inductance at the connecting
position~\cite{Grajcar05,Grajcar06}.

\section{Non-perturbative dispersive qubit readout}
\begin{figure*}[tbp]
    \centering
    \includegraphics[width=18cm]{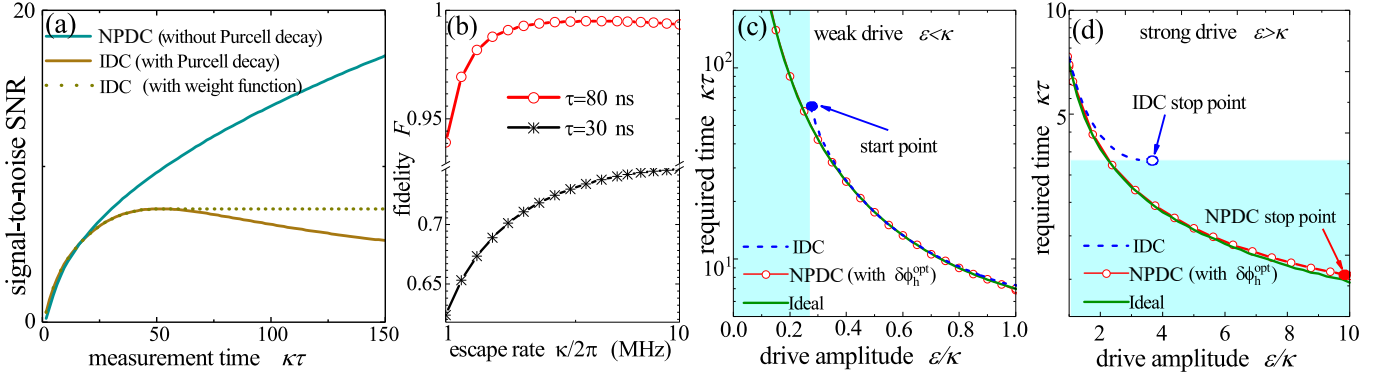}
    \caption{(a) Signal-to-noise ratio for the NPDC- and IDC-based readout mechanisms
        versus the integrated measurement time $\kappa\tau$. (b) Measurement
        fidelity $F_{s}$ versus photon escape rate $\kappa$ at
        $\tau=30~\text{ns}$ and $\tau=80~\text{ns}$. In (a) and (b) we set
        $\epsilon=\chi_{z}$. (c, d) The measurement time $\kappa\tau$
        required to reach the fidelity $99.99\%$ versus the drive
        amplitude $\epsilon$ for: (c) weak ($\epsilon/\kappa<1$) and (d)
        strong ($\epsilon/\kappa\geq 1$) drives. In the IDC readout, the
        solid (empty) circle marks the start (stop) point, indicating the
        lower (upper) bound of $\epsilon$. In the NPDC readout, we employ the
        optimal shifted homodyne angle $\delta\phi_{h}^{\text{opt}}$ for
        different drive strengths, as discussed in Appendix~E. The red solid
        point is the NPDC stop point due to the nonlinear effects.
        We set the parameters as: $\lambda=0.1$,
        $K/(2\pi)=100~\text{kHz}$, $\chi_{z}/(2\pi)= 8~\text{MHz}$, and
        $\kappa/(2\pi)=16~\text{MHz}$. The cyan areas are the regimes that cannot
        be reached in the IDC-based readout mechanism due to the Purcell effects.}
    \label{fig4m}
\end{figure*}
Based on the layout in Fig.~\ref{fig1m}, one can realize an ideal
QND readout of the flux qubit via the coupling Hamiltonian
$H_{\alpha}$ without being disturbed by the Purcell effects. To
compare the IDC and NPDC readouts of the qubit, below we assume
\begin{equation}
\chi_{z}^{I}=\chi_{z}^{D}=\chi_{z}, \quad K_{D}=K_{I}=K.
\end{equation}
Applying an incident field $a_{\text{in}}$ in the left port of the
resonator at the resonator frequency $\omega_{r}'$, the quantum
nonlinear Langevin equation for the resonator operator reads
\begin{equation}
\frac{da(t)}{dt}=-i\chi_{z}\sigma_{z} a(t)-2iK\langle N(t)\rangle
a(t)-\frac{1}{2}\kappa a(t) -\sqrt{\kappa}a_{\text{in}}(t).
\label{langevin}
\end{equation}
For the IDC readout, $\langle N(t)\rangle=\langle
n(t)\rangle\sigma_{z}$ is due to the qubit-dependent Kerr
nonlinearity, where $\langle n(t)\rangle=\langle
a^{\dag}(t)a(t)\rangle$ is the average intracavity photon number.
For the NPDC-based readout mechanism, $\langle N(t)\rangle=\langle
n(t)\rangle$ results from the standard Kerr term. This input field
$a_{\text{in}}(t)=\alpha_{\text{in}}+d_{\text{in}}(t)$  is
characterized by its mean value (a coherent drive)
$\alpha_{\text{in}}=-\epsilon e^{i\theta_{d}}/\sqrt{\kappa}$ and
fluctuation $d_{\text{in}}(t)$. Due to the dispersive coupling,
the qubit state is encoded in the output quadrature
$Y(\phi_{h})=a_{\text{out}}^{\dag}e^{i\phi_{h}}+a_{\text{out}}
e^{-i\phi_{h}}$. The measurement corresponds to a homodyne
detection of $Y(\phi_{h})$ with an integration time $\tau$, i.e.,
\begin{equation}
M(\tau) =\sqrt{\kappa} \int_{0}^{\tau} Y(\phi_{h}) dt.
\end{equation}
We first consider an ideal readout with $K=0$. By formally integrating
Eq.~(\ref{langevin}) and using the input-output relation
$a_{\text{out}}=\sqrt{\kappa} a+a_{\text{in}}$, we obtain the
separation signal $M_{s}=\langle M_{s}\rangle
_{|e\rangle}-M_{s,|g\rangle}$ (with $\langle \sigma_{z}\rangle
=\pm1$) as
\begin{eqnarray}
&&M_{s}(\tau)=4\epsilon \sin{2\theta_{q}}\sin(\theta_{d}-\phi_{h}) \notag  \\
&&\times\left[\tau-\frac{4\cos^{2}(\theta_{q})}{\kappa}\left(1-\frac{\sin(\chi_{z}\tau+2\theta_{q})}{\sin(2\theta_{q})}e^{-\frac{1}{2}\kappa
    \tau}\right)\right], \label{signalS}
\end{eqnarray}
where $\theta_{q}=\arctan({2\chi_{z}/\kappa})$ is the rotating
angle of the output field. We integrated the Langevin
equation in (\ref{directnn}), to obtain Eq.~(\ref{alphabar}),
which shows the coherent amplitudes $\alpha$ of the intracavity
field. In our derivation, we approximately used the classical part
of the intracavity field operator to describe the photon number,
i.e., $\langle n(t) \rangle=|\alpha(t)|^{2}$. As derived in
Appendix~E, $\langle n(t) \rangle$ is given by
\begin{eqnarray}
\langle n(t) \rangle&=&\left(\frac{2\epsilon}{\kappa}\right)^{2}\cos(\theta_{q})\bigg[1+\exp(-\kappa t) \notag \\
&-&2\cos(\chi_{z}\langle\sigma_{z}\rangle t)\exp\left(-\frac{1}{2}\kappa t\right)\bigg]. \label{intraphoton}
\end{eqnarray}
In the steady state ($\kappa t\gg 1$), the intracavity photon number is $n\backsimeq (2\epsilon/\kappa)^2\cos(\theta_{q})$.
The fluctuation $d_{\text{out}}(t)$
introduces noise into the measurement signal. For the vacuum
input, the noise reads~\cite{Didier152}
\begin{equation}
M_{N}^{2}(\tau)=[\langle
M_{N}^{2}(\tau)\rangle _{|e\rangle}+\langle M_{N}^{2}(\tau)\rangle
_{|g\rangle}]=2\kappa \tau. \label{noisee}
\end{equation}

In the long-time
limit $\kappa \tau\gg 1$, the signal-to-noise-ratio
$\mathcal{R}=M_{s}(\tau)/M_{N}(\tau)$ is optimized by setting
$\theta_{d}-\phi_{h}=\pi/2$ and $\theta_{q}=\pi/4$ (i.e.,
$\chi_{z}=\kappa/2$). The measurement fidelity is defined as
\begin{equation}
F_{m}=\frac{1+\text{erf}(\mathcal{R}/2)}{2}, \label{fidelity}
\end{equation}
where $\text{erf}(x)$ is the error function.

As discussed in Appendix~E, the effects of the nonlinearities in
the IDC and NPDC readouts are different: the Kerr nonlinearity in
the IDC readout is qubit-dependent, and symmetrically reduces the
effective cavity pull~\cite{Boissonneault08}, which causes a poor
signal separation if $\langle a^{\dag}(t)a(t)\rangle$ is large.
For the NPDC readout, $K$ leads to asymmetric rotation angles of
the cavity field in the phase space. However, the
signal-separation distance is still high, even for large $\langle
a^{\dag}(t)a(t) \rangle$. Moreover, for the IDC-based readout
mechanism, because $[H_{x},\sigma_{z}]\neq0$, $H_{\text{d}}$ is
not an ideal QND readout Hamiltonian. There is a qubit Purcell
decay channel $\Gamma_{p}$ via the readout resonator, which is
proportional to the photon escape rate $\kappa$. Assuming the
qubit relaxation is limited by the Purcell decay, the readout $\mathcal{R}$
can be numerically derived by replacing $\sigma_{z}$ by
\begin{equation}
\langle \sigma_{z} \rangle
(t)=[1+\langle \sigma_{z} \rangle (0)]\exp(-\Gamma_{p}t)-1
\end{equation}
in Eq.~(\ref{langevin}). It is hard to obtain analytical results
of Eq.~(\ref{langevin}) by including both $K$ and $\Gamma_{p}$.
Thus, below we present only numerical
results~\cite{Johansson12qutip,Johansson13qutip}. We first plot
the signal-to-noise-ratio versus time in Fig.~\ref{fig4m}(a). In the IDC readout, due
to the Purcell decay, the signal-to-noise-ratio decreases after reaching its
maximum. In our simulations, the separation signal $M_{s}$ indeed
saturates at a constant level when increasing the measurement
time. However, in Fig.~\ref{fig4m}(a), we plot the signal-to-noise-ratio, which is
defined as $\mathcal{R}=M_{s}/\sqrt{2\kappa \tau}$. By increasing the
measurement time $\tau$, the signal separation $M_{s}$ finally
reaches its steady value, while a homodyne detector continuously
collects the input noise. Therefore, in our discussions, $\mathcal{R}$
decreases with time as shown in Fig.~\ref{fig4m}(a). Note that our
definition is different from that in the experimental
works~\cite{Gambetta2007, Bultink2018, Heinsoo2018}. The
weight function $W(t)$ determines how much of the signal power is
integrated at time $t$. This function $W(t)$ can, in principle, be
chosen a square (boxcar) function or can be optimized based on
certain experimental implementations~\cite{Bultink2018}. In
Fig.~\ref{fig4m}(a), we use a simple square function, which leads
to a constant value of $\mathcal{R}$ after reaching its highest point.

For the NPDC readout, the escaping photon does not lead to the
decay of the qubit states, and $\mathcal{R}$ is proportional to
$\sqrt{2\kappa \tau}$ in the long-time limit $\kappa
\tau\gg1$~\cite{Didier15,Didier152}. One may try to suppress
$\Gamma_{p}$ by reducing $\kappa$. However, to achieve a fast
qubit readout, $\kappa$ should be large enough to allow readout
photons to escape quickly.
The relation can be clearly found in Fig.~\ref{fig4m}(b): 
for a certain integrated time $\tau$, the readout fidelity
increases with $\kappa$. Therefore, to reduce the Purcell decay,
one should decrease the measurement speed with a relatively low
$\kappa$ in the IDC readout. However, this trade-off relation does
not exist in the NPDC readout: it is without dipole-field
coupling, and the qubit QND readout is not disturbed by the
Purcell decay. One can employ a large $\kappa$ to speed up the
readout.
\begin{figure}[tbp]
    \centering
    \includegraphics[width=8.9cm]{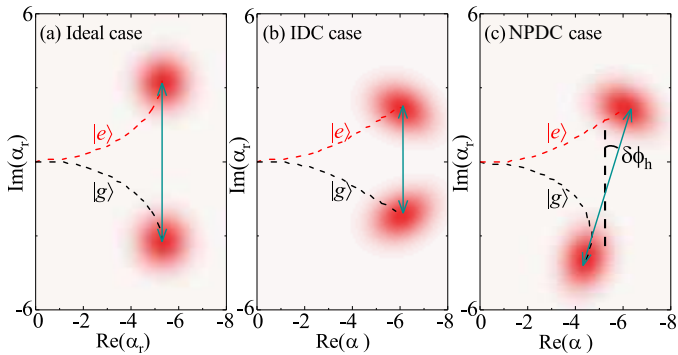}
    \caption{The Wigner distributions of the intracavity field
        for (a) the ideal, (b) IDC, and (c) NPDC readouts for the same measurement time $\kappa \tau=3$.
        The red (black) curves are the time-dependent evolution trajectories in phase space, and the upper (lower) signals correspond to the qubit being in its excited (ground) state.
        The arrows represent signal separation distances and directions.
        The parameters adopted here are the same as those in Fig.~\ref{fig4m}(d),
        and the drive strength is assumed to be the same as for the stop point.
        For the NPDC readout in (a), the signal separation direction is rotated by an angle $\delta\phi_{h}$.} \label{fig5m}
\end{figure}

As shown in Fig.~\ref{fig5m} (with the same
measurement time $\kappa \tau=3$), one can find that the Wigner functions for the IDC- and NPDC-based
readout mechanisms are not perfectly symmetric Gaussian functions.
This is due to the Kerr nonlinearity, which modifies the coherent
state of the resonator field. Consequently, the collected noise in
the homodyne measurement for a given quadrature direction in phase
space can be different from the ideal result given in
Eq.~(\ref{noisee}) assuming a classical resonator field.
Nevertheless, we employ the Heisenberg-Langevin equation and treat
the resonator field to be classical in Fig.~\ref{fig4m}. Precise
numerical calculations of the evolution of our system, including
the quantized resonator field with many injected photons, require
to assume the Hilbert-space dimension of the order of about
hundreds and, thus, require time-consuming numerical calculations.
However, for the parameters considered in our manuscript, treating
the resonator field to be coherent leads only to very small
difference, compared to the fully quantum treatment of the entire
system.

\begin{figure}[tbp]
    \centering
    \includegraphics[width=8.7cm]{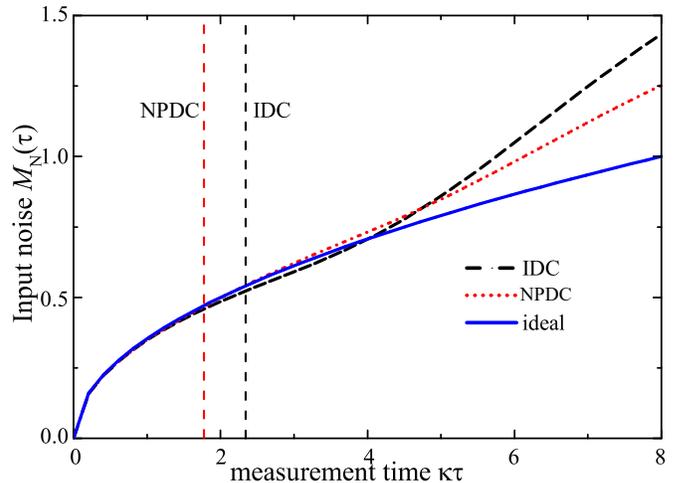}
\caption{The measurement noise $M_{N}(\tau)$ for the ideal, IDC,
and NPDC readouts obtained by quantum simulations. The dashed
lines correspond to the required measurement time for the IDC and
NPDC proposals. Parameters assumed here are the same as those for
the stop point of the IDC readout shown in Fig.~\ref{fig4m}(d).}
\label{fig6m}
\end{figure}

To verify this classical approximation of the resonator field, we
plotted Fig.~\ref{fig6m} by applying the quantum treatment of the
resonator field for the parameters at the IDC stop point in
Fig.~\ref{fig4m}(d). This figure shows the effect of the quantum
noise $M_{N}(\tau)$ changing with the measurement time for the
ideal, NPDC, and IDC readouts. The vertical lines correspond to
the required time for the IDC and NPDC readouts in
Fig.~\ref{fig4m}(d), respectively, where the noise shift, which is
induced by the Kerr effect, is negligible compared to the ideal
readout and one can employ our analytical results given in
Eq.~(\ref{noisee}) to calculate $\mathcal{R}$. If one insists on
achieving a higher fidelity, the required time becomes longer and,
thus, one should calculate the input quantum noise changing with
the measurement time for different methods in the fully quantum
treatment of the resonator field. Moreover, by comparing the IDC
and NPDC results in the long-measurement-time limit
($\kappa\tau\gg 1$), we find that the noise for the NPDC readout
increases more slowly compared to the noise in the IDC readout.
This is another advantage of our proposal.

In Figs.~\ref{fig4m}(c) and \ref{fig4m}(d), we plot the time
$\kappa \tau$ required to reach the fidelity $99.99\%$ as a
function of the drive strength $\epsilon$. Figure~\ref{fig4m}(c)
corresponds to the weak-drive limit ($\epsilon<\kappa$). Due to
the qubit Purcell decay, there is a lower limit of
$\epsilon_{\text{min}}$ (start point) for the drive amplitude in
the IDC readout. If $\epsilon<\epsilon_{\text{min}}$ (cyan area),
the measurement can never reach the desired fidelity even if
taking an infinitely-long time. However, for the NPDC readout (red
dotted curve) based on our proposal, the ideal fidelity can be
reached in principle for $\epsilon<\epsilon_{\text{min}}$.

We should recall that for the NPDC-based readout mechanism, we
cannot increase the photon number due to some limitations of
Eq.~(\ref{Ncrt}). In Fig.~\ref{fig4m}(d), we marked the
corresponding stop point (due to $n_{c}$) for the NPDC readout
(red solid point), which is around $\epsilon/\kappa\simeq 10$ and
far from the stop point of the IDC readout. Note that the critical
photon number $n_{c}$ decreases when the flux bias
$\Phi_{\text{ext}}$ is close to $\Phi_{0}/2$. As shown in
Fig.~\ref{fig2m}, there is a trade-off relation between the flux
sensitivity $R$ (i.e., the NPDC coupling strength) and $n_{c}$,
which might be one of the obstacles for achieving much shorter
readout times. In future studies, to reach the minimum readout
time, one can optimize the parameters of the whole readout
circuits.

In the strong-drive limit, $\epsilon>\kappa$
[Fig.~\ref{fig4m}(d)], for both two readout mechanisms, the
required time $\kappa \tau$ is significantly reduced.
Unfortunately, the IDC-based readout mechanism encounters
another two Purcell limitations: First, the effective cavity pull
is significantly reduced as
\begin{equation}
|\xi_{I}|=\left|\chi_{z}\left(1- \frac{\langle
    n(t) \rangle}{2n_{c} } \right)\right|,  \\
\label{cavitypullabs}
\end{equation}
which leads to a reduction of the signal separation. By comparing
Fig.~\ref{fig5m}(a) and (b), it can be found that, for the IDC readout, the signal
separation distance (oliver arrows) in phase space is much smaller
than that in the ideal readout. Consequently, the required time
becomes much longer than that for the ideal readout. Second, to
avoid photon-induced qubit-error transitions, the intracavity
photon number should be much smaller than the critical photon
number $n_{c}=1/(4\lambda^{2})$. This sets another upper bound
limitation $\epsilon_{\text{max}}=\kappa/(2\sqrt{2}\lambda)$ for
the drive strength~\cite{Blais04}. The measurement time $\tau$
cannot be shortened below the stop point.

For large $\langle n(t) \rangle$, the Kerr nonlinearity in the
NPDC-based readout mechanism also induce apparent effects.
However, as shown in Fig.~\ref{fig5m}, its main effect is to
change the signal separation direction with a small angle $\delta
\phi_{h}$~\cite{Eichler2014}, while the signal separation distance
is still large compared with the IDC readout. Moreover,
intracavity photons do not cause qubit-state error flips and,
therefore, in principle, there is no such a stop point due to the
Purcell effects. To minimize the Kerr effects, we can slightly
shift the homodyne angle $\phi_{h}$ by an optimized small angle
$\delta\phi_{h}^{\text{opt}}$. The detailed method about how to
shift the measurement angle can be found in Appendix~E. As seen in
Fig.~\ref{fig4m}(d), the required time can be quite close to the
ideal readout. Therefore, by injecting many photons, the
measurement time can go far below the Purcell-effect regime.

\section{Discussions}
\subsection{Multiqubit readout via a single resonator}
It is also possible to employ an array of SQUIDs to terminate the
measurement resonator (see Ref.~\cite{Sandberg2008}). As discussed
in Appendix~B, the effective nonlinear inductance of each SQUID
can be tuned independently via the flux $\Phi_{\text{ext},j}$
produced by an individual flux-bias line. Considering the $j$th
SQUID interacting with the $j$th flux qubit ($\sigma_{z,j}$) via
its circulating current $I_{\text{cir},\alpha j}$, the interaction
Hamiltonian is
\begin{equation}
H_{\alpha,\Sigma}=\sum_{j} \chi_{z,i}\sigma_{z,j}a^{\dag}a, \quad \chi_{z,i}=R(\Phi_{\text{ext},j})I_{\text{cir},\alpha j}M_{j},
\end{equation}
with $M_{j}$ being the mutual inductance. To readout the
$m$th qubit without being disturbed by other qubit-resonator
couplings, one can tune $R(\Phi_{\text{ext},j})$ to zero with
$\Phi_{\text{ext},j}=0$ for $j\neq m$, while keeping
$R(\Phi_{\text{ext},m})$ around point B (see Fig.~\ref{fig2m}). In
this case, the resonator is employed as a shared readout resonator
for each individual qubit. Moreover, we could realize a joint
readout of multiqubit states~\cite{Majer2007}. For the example of two
qubits, we set $\chi_{z,1}/2=\chi_{z,2}=\chi_{0}$. The two-qubit basis
$\{ |e,e\rangle, |e,g\rangle, |g,e\rangle, |g,g\rangle \}$
corresponds to four different rotation angles (in phase space)
$\theta_{qN}=\arctan({N\chi_{0}/\kappa})$, with $N=\pm1, \pm3$ for
the output field, which represents four separated pointer states.
This multi-SQUID layout enables scalability for an ideal
qubit-joint QND readout.

By assuming similar values of the flux sensitivity
$R(\Phi_{\text{ext},j})$ of all the SQUIDs (labeled by the index
$j$), the phase drops across each SQUID are also
similar~\cite{Eichler2014}. Due to this property, the Kerr
nonlinearity $K'_{D}$ increases linearly with the SQUID number
$n$, i.e., $K'_{D}=nK_{D}$. In all our discussions, the Kerr
nonlinearity always decreases the readout fidelity. Therefore,
this scalable proposal works well when considering only several
SQUIDs and a weak drive field. Beyond these regimes, we should
find better proposals for such a qubit-QND readout.

\subsection{Dynamical range of the SQUID-terminated resonator}

Finally, we want to discuss the dynamical range of the measurement
SQUID-terminated resonator. The differential
equation~(\ref{langevin}) for NPDC readout is nonlinear, and the
steady-state solution for the intracavity-photon number can be
solved from a cubic equation. To analyze this problem, we need to
define the dimensionless effective detuning as
\begin{equation}
\delta^{s}=\frac{\chi_{z}^{D}}{\kappa}\langle \sigma_{z}\rangle, \qquad
\delta^{s}\in \left[-\left|  \frac{\chi_{z}^{D}}{\kappa} \right|,  \left| \frac{\chi_{z}^{D}}{\kappa}\right|  \right].
\end{equation}
As derived in Ref.~\cite{Eichler2014}, the critical detuning is $\delta^{s}_{\text{crt}}=-\sqrt{3}/2$, below which the cubic equation might have three solutions for the intracavity-field intensity. Both, the smallest and the largest solutions, are stable for the whole system. However, the intermediate one is unstable, around which the field in the readout resonator might bifurcate. During the dispersive readout, better for the system to avoid this highly nonlinear regime.

As shown Fig.~\ref{fig4m}(b), a rapid photon escaping rate $\kappa$ improves the readout fidelity. In an experimental implementation, by adopting a large $\kappa$, the dimensionless detuning $|\delta^{s}|$ is a small parameter. For example, in Figs.~\ref{fig4m}(c) and (d), we adopt $\delta^{s}\in[-1/2, 1/2]$, which is out of the bistable regime. Therefore, our proposal can effectively avoid bistability problems induced by the Kerr nonlinearity.
\section{Conclusions}

We showed how to realize an ideal QND readout of a flux qubit via
its non-perturbative dispersive coupling with a SQUID-terminated measurement
qubit. The coupling can be conveniently switched on and off via an
external flux control. Compared with the conventional induced
dispersive coupling based on the Rabi model, this mechanism is
free of dipole-field interactions and, therefore, it is not
deteriorated by the Purcell effects. We can employ a strong drive
field and a quick photon escape rate. Thus, both measurement
fidelity and speed can avoid the Purcell limitations. Considering
a single resonator, which is terminated by a series of SQUIDs,
this proposal is scalable and tunable to realize a multi-qubit
joint QND readout. In future studies, this proposed method might
be developed to include a weak continuous measurement to monitor
the superconducting flux qubit~\cite{Groen13,Tan2015}. Moreover,
this method can be applied to other weak-signal measurements, such
as detecting virtual photons or qubit-excited states in the
ultrastrong light-matter coupling
regime~\cite{Cirio17,Kockum2018}.

\section{acknowledgments}
The authors acknowledge fruitful discussions with Xiu Gu, Yu-Xi
Liu, Zhi-Rong Lin, and Wei Qin. X.W. thanks Yun-Long Wang for
helping to depict the schematic diagrams. X.W. is supported by
China Postdoctoral Science Foundation No. 2018M631136, and the
Natural Science Foundation of China under Grant No. 11804270. A.M.
and F.N. acknowledge the support of a grant from the John
Templeton Foundation. F.N. is supported in part by the: MURI
Center for Dynamic Magneto-Optics via the Air Force Office of
Scientific Research (AFOSR) (FA9550-14-1-0040), Army Research
Office (ARO) (Grant No. Grant No. W911NF-18-1-0358), Asian Office
of Aerospace Research and Development (AOARD) (Grant No.
FA2386-18-1-4045), Japan Science and Technology Agency (JST) (via
the Q-LEAP program, and the CREST Grant No. JPMJCR1676), Japan
Society for the Promotion of Science (JSPS) (JSPS-RFBR Grant No.
17-52-50023, and JSPS-FWO Grant No. VS.059.18N), and the
RIKEN-AIST Challenge Research Fund.

Note added: After posting the e-print of this work in the
arXiv~\cite{Wang2018x}, the work by Dassonneville \emph{et al.} on
the same topic was posted in the arXiv~\cite{Dasson2019}. That
paper describes a protocol very similar to ours, but for a
different type of a superconducting qubit, which can be viewed as
an experimental realization of the QND measurements of
superconducting qubit via non-perturbative dispersive coupling.  Those
studies indicate that the concept of a QND readout via the NPDC
mechanism is experimentally feasible and can receive much more
attention, both theoretical and experimental, in the near future.

\section*{APPENDICES}

\setcounter{equation}{0} \renewcommand{\theequation}{A\arabic{equation}}
\setcounter{figure}{0} \renewcommand{\thefigure}{A\arabic{figure}}\

\begin{appendix}

\section{Induced dispersive coupling and Purcell decay}

In a typical system based on superconducting quantum circuits, the
conventional light-matter dispersive coupling is based on
dipole-field interactions. In the large-detuning regime $g_{x}\ll
\Delta_{d}=\omega_{q}-\omega_{a}$ (where $\omega_{q}$ and
$\omega_{r}$ are the qubit and resonator frequency, respectively),
the system Hamiltonian is approximately described by the
Jaynes-Cummings (JC) Hamiltonian
\begin{equation}
H_{R0}=\frac{\omega_{q}}{2}\sigma_{z}+\omega_{r}a^{\dag}a+g_{x}(a\sigma_{+}+a^{\dag}\sigma_{-}).
\label{Hrabi}
\end{equation}

In a qubit dispersive readout, one often injects many photons into
the resonator to speed up such measurement. Once the photon number
is large, it is necessary to push the dispersive coupling into
higher-order nonlinear terms. Here we follow the approaches in
Refs.~\cite{Boissonneault08,Boissonneault09}, and derive a more
exact nonlinear dispersive coupling Hamiltonian. We first define
the unitary transformation $U_{D}$ as~\cite{Boissonneault09}:
\begin{eqnarray}
U_{D}&=&\exp\left\{-\Theta(N_{J})(a^{\dag}\sigma_{-}-a\sigma_{+})\right\}, \qquad \notag \\
\Theta(N_{J})&=&-\frac{\arctan(2\lambda
    \sqrt{N_{J}})}{2\sqrt{N_{J}}},
\end{eqnarray}
where
$$\lambda=g_{x}/(\Delta_{d}),\quad \Delta_{d}=\omega_{q}-\omega_{r},$$
and
$$N_{J}=a^{\dag}a+|e\rangle\langle e|$$ is a function of the total
excitation number operator of the system. Applying the
transformation $U_{D}$ to the JC Hamiltonian $H_{R0}$, the
off-diagonal terms can be eliminated, and yielding
\begin{equation}
H_{R0}=\frac{\omega_{q}}{2}\sigma_{z}+\omega_{r}a^{\dag}a-\frac{\Delta_{d}}{2}\left(1-\sqrt{1+4\lambda^{2}N_{J}}\right)\sigma_{z}.
\label{Hdis}
\end{equation}
This equation is still the exact diagonalized solution for the
system Hamiltonian without any approximation. To obtain the
dispersive coupling, we can expand
$\sqrt{1+4\lambda^{2}N_{J}}$ to second order in
$4\lambda^{2}N_{J}$ to find~\cite{Boissonneault09}
\begin{eqnarray}
H_{R0}&=&\omega_{r}'a^{\dag}a+\frac{\omega_{q}}{2}\sigma_{z} \notag \\
&&+\chi_{z}^{I} a^{\dag}a\sigma_{z}+K_{I}(a^{\dag}a^{\dag}aa)\sigma_{z},
\label{inducedis}
\end{eqnarray}
where $\omega_{r}'$ is the shifted resonator frequency,
$$\chi_{z}^{I}=g_{x}^{2}(1-2\lambda^{2})/\Delta_{d} \backsimeq
g_{x}^{2}/\Delta_{d}$$ is the IDC strength, and
$$K_{I}=-g_{x}^{4}/\Delta_{d}^{3}$$
is the qubit-dependent Kerr nonlinearity strength. Note that the
validation of this perturbation result requires that
Eq.~(\ref{inducedis}) does not only depend on a small parameter
$\lambda\ll 1$, but also requires that the total excitation number
satisfies $\langle N_{J} \rangle \ll 1/(4\lambda^{2})$, which
results in a critical photon
$n_{c}=1/(4\lambda^{2})$~\cite{Blais04}. In a qubit measurement,
the intracavity photon number should be much smaller than $n_{c}$.

The coupling Hamiltonian between the measurement and the
environment is
$$H_{\kappa}=\int_{0}^{\infty}\sqrt{\kappa(\omega)}[b^{\dag}(\omega)+b(\omega)](a+a^{\dag})d\omega,$$
where $b(\omega)$ is the annihilation operator of the
environmental mode $\omega$. Applying the unitary transformation
$U_{D}$ to the field operators $(a+a^{\dag})$, we obtain
\begin{equation}
U_{D}^{\dag} (a+a^{\dag})U_{D}\simeq
(a+a^{\dag})+\lambda(\sigma_{-}+\sigma_{+})+O'(\lambda^{2}).
\label{dampo}
\end{equation}
One can find that the field operator acquires an extra part
related to the qubit operators $\sigma_{\pm}$ in the dressed
basis. In the interaction picture and applying the rotating-wave
approximation to the Hamiltonian $H_{\kappa}$, we obtain
\begin{eqnarray}
H_{\kappa}&=&\int_{0}^{\infty}  d\omega
\sqrt{\kappa(\omega)}[b(\omega)a^{\dag}e^{-i(\omega-\omega_{r})t}+\text{H.c.}] \notag \\
\!&+&\!
\lambda\!\int_{0}^{\infty} d\omega
\sqrt{\kappa(\omega)}[b(\omega)\sigma_{+}e^{-i(\omega-\omega_{q})t}+\text{H.c.}],
\label{sec}
\end{eqnarray}
where the last term describes an additional Purcell decay channel
for the qubit. The cavity is assumed to couple with a thermal
environment with zero average boson number. Following the
standard steps of deriving the master equation, we find that the
last term adds an extra qubit decay with rate
$\Gamma_{p}=\lambda^{2}\kappa(\omega_{q})$. In this work,
we assume that $\kappa(\omega_{q})$ is not frequency
dependent and equals to the photon escape rate $\kappa$.

\setcounter{equation}{0} \renewcommand{\theequation}{B\arabic{equation}}

\section{SQUID-terminated transmission line resonator}

\subsection*{B.1 Tuning the resonator frequency via SQUID: Linear approximation}
As shown in Fig.~\ref{fig1m}, we consider a transmission
line resonator (TLR) (along the $x$ axis with length $d_{0}$)
short-circuited to ground by terminating its right side with
a dc SQUID (at the position $x=d_{0}$)~\cite{Wallquist06,Johansson09L}. The two Josephson
junctions of the SQUID are assumed to be symmetric with identical
Josephson energy $E_{s0}$ and capacitance $C_{s}$. The effective
Josephson energy of the SQUID is tuned with the external flux
$\Phi_{\text{ext}}$, according to the relation
$E_{s}=2E_{s0}|\cos(\pi \Phi_{\text{ext}}/\Phi_{0})|$ ($\Phi_{0}$ is
the flux quantum). For an asymmetric SQUID with two different Josephson
energies $E_{s1}$ and $E_{s2}$, the effective Josephson energy of the SQUID is given by
\begin{equation}
E'_{s}=E_{\Sigma}\sqrt{\cos^{2}\left(\frac{\pi \Phi_{\text{ext}}}{\Phi_{0}}\right)+d_{0}\sin^{2}\left(\frac{\pi \Phi_{\text{ext}}}{\Phi_{0}}\right)},
\label{asyme}
\end{equation}
where
\begin{equation}
d_{0}=\frac{E_{s1}-E_{s2}}{E_{s1}+E_{s2}},
\end{equation}
is the junction asymmetric parameter. From Eq.~(\ref{asyme}), the asymmetric effects can be neglected under the condition
\begin{equation}
d_{0}<\cot\left(\frac{\pi \Phi_{\text{ext}}}{\Phi_{0}}\right).
\label{d0a}
\end{equation}
In our work, we consider $\Phi_{\text{ext}}/\Phi_{0}\in[0.46,0.48]$, which results $d_{0}<7\%$. The present fabrication technology can control the junction asymmetric parameter within $d_{0}< 2\%$. Therefore, the condition in Eq.~(\ref{d0a}) is within reach of current experiments~\cite{Gu2017}.

Note that the SQUID has a nonlinear inductance
$$L_{s}(\Phi_{\text{ext}})=\frac{\Phi_{0}^{2}}{(2\pi)^2E_{s}(\Phi_{\text{ext}})}$$
and its Lagrangian is written as~\cite{Sandberg2008,Wallquist06}
\begin{equation}
\mathcal{L}=\sum_{i=1,2}\left(\frac{\Phi_{0}^{2}C_{s}}{2(2\pi)^2}\dot{\phi_{i}}^{2}+E_{s0}\cos{\phi_{i}}\right).
\label{LSQUID1}
\end{equation}
Setting $\phi_{+}=(\phi_{1}+\phi_{2})/2$ and
$\phi_{-}=(\phi_{1}-\phi_{2})/2=\pi \Phi_{\text{ext}}/\Phi_{0}$,
we rewrite Eq.~(\ref{LSQUID1}) as
\begin{equation}
\mathcal{L}=\frac{\Phi_{0}^{2}C_{s}}{(2\pi)^2}\dot{\phi}_{+}^{2}+E_{s}\cos{\phi_{+}}.
\label{LSQUID}
\end{equation}
Given that $E_{s}\gg (2e)^2/(2C)$, the zero-point fluctuation in
the plasma oscillation is of small amplitude with
$\phi_{z0}=\sqrt[4]{4e^{2}/(2C_{s}E_{s})}\ll 1$, the SQUID is
around its quantum ground state~\cite{Johansson14}. The
SQUID can be seen as a harmonic oscillator with Lagrangian~\cite{Wallquist06}
\begin{equation}
\mathcal{L}\simeq
\frac{\Phi_{0}^{2}C_{s}}{(2\pi)^2}\dot{\phi}_{+}^{2}-\frac{E_{s}}{2}\phi_{+}^{2}.
\label{LSQUIDa}
\end{equation}
Let us denote the transmission-line capacitance and inductance per
unit length as $C_{0}$ and $L_{0}$, respectively. The dynamics
of the field along the transmission-line direction (denoted as the
$x$ axis) is described by the Helmholtz wave equation
\begin{equation}
\frac{\partial^{2} \psi(x,t)}{\partial
    t^{2}}-v^{2}\frac{\partial^{2} \psi(x,t)}{\partial x^{2}}=0,
\label{waveeq}
\end{equation}
where $v=1/\sqrt{L_{0}C_{0}}$ is the wave velocity. At $x=0$ with
a large capacitance $C_{g}$, the bound condition is $\partial_{x}
\psi(0,t)=0,$ which requires that the wavefunction solutions of
Eq.~(\ref{waveeq}) for a mode $k$ have the form
$\psi(x,t)=\psi_{0}\sin(kvt)\cos(kx)$. At $x=d_{0}$, the boundary
conditions are~\cite{Wallquist06,Johansson09}:

\begin{gather}
\psi(d_{0},t)=\phi_{+}(t),     \notag   \\
2C_{s}\ddot{\psi}(d_{0},t)+\frac{(2\pi)^2}{\Phi_{0}^{2}}E_{s}\psi
(d_{0},t)+\frac{\partial_{x} \psi(d_{0},t)}{L_{0}}=0.
\label{bound1}
\end{gather}
By substituting the wave function into Eq.~(\ref{bound1}), one can
find that the mode frequency $\omega_{r}=vk$ of the resonator can
be derived from the following transcendental
equation~\cite{Pogorzalek17}:
\begin{equation}
\frac{\pi\omega_{r}}{2\omega_{0}}\tan\!\!\bigg(\frac{\pi\omega_{r}}{2\omega_{0}}\bigg)=\frac{(2\pi)^2}{\Phi_{0}^{2}}L_{t}E_{s}(\Phi_{\text{ext}})-\frac{2C_{s}}{C_{t}}\bigg(\frac{\pi\omega_{r}}{2\omega_{0}}\bigg)^{2},
\label{diseq}
\end{equation}
where $L_{t}=d_{0}L_{0}$ and $C_{t}=d_{0}C_{0}$ are the total
inductance and capacitance of the resonator, respectively. The
fundamental frequency of the quarter-wavelength resonator is
$\omega_{0}=\pi v/(2d_{0})$. By assuming that the capacitances of
the Josephson junctions $C_{s}$ are much smaller compared with the
total capacitance $C_{t}$, we neglect the last term in
Eq.~(\ref{diseq}). Because the total inductance $L_{t}$ strongly
exceeds that of the SQUID nonlinear inductance
$L_{s}(\Phi_{\text{ext}})$, we find $\omega_r/\omega_{0}\simeq1$,
and rewrite Eq.~(\ref{diseq}) as
\begin{equation}
\left[\frac{\pi\omega_{r}}{2\omega_{0}}\tan\bigg(\frac{\pi\omega_{r}}{2\omega_{0}}\bigg)\right]^{-1}=\frac{L_{s}(\Phi_{\text{ext}})}{L_{t}}.
\label{fstorder}
\end{equation}
By expanding the left-hand side of Eq.~(\ref{fstorder}) with
$(\pi\omega_r)/(2\omega_{0})$ around $\pi/2$ to first order,
we obtain
\begin{equation}
\omega_{r}=\omega_{0}\left[1-\frac{L_{s}(\Phi_{\text{ext}})}{L_{t}}\right].
\label{nonlinfre}
\end{equation}
From this equation, we find that the external flux
$\Phi_{\text{ext}}$ through the SQUID determines its nonlinear
inductance, which eventually shifts the mode frequency
$\omega_{r}$. Similar to the discussions in Ref.~\cite{Johansson09},
this parametric bound condition changes the resonator effective
length only slightly, which is akin to a moving mirror for
modulating the effective wavelength in the optomechanical system.
We assume that the external flux is composed of a prebiased
static part $\Phi_{\text{ext}}^{0}$ and a small deviation part
$\delta \Phi_{\text{ext}}\ll \Phi_{\text{ext}}^{0}$, and  write
the mode frequency as
\begin{equation}
\omega_{r}=\omega_{r0}+\frac{\partial \omega_{r}}{\partial
    \Phi_{\text{ext}}}\Big|_{\Phi_{\text{ext}}^{0}}\delta
\Phi_{\text{ext}},
\end{equation}
where the shifted mode frequency $\omega_{r0}$ and its flux
sensitivity $R$ are expressed in Eq.~(\ref{Randfre}).
Note that in our discussions we assume that the dc-SQUID loop
inductance can be neglected when compared with $L_{s}$, which can
be easily satisfied in experiments~\cite{Eichler18}. Therefore,
the frequency jump effects of the mode frequency due to its
hysteretic flux response can also be
neglected~\cite{Pogorzalek17}.

\subsection*{B.2 Resonator self-Kerr nonlinearity}

Because the SQUID is a nonlinear element, attaching it at the end
of the resonator makes the \emph{entire} system nonlinear. Here we
want to estimate the amount of such nonlinearity. In
Eq.~(\ref{LSQUIDa}), we approximately viewed the SQUID as a linear
circuit element by neglecting the higher-order terms. To obtain
the nonlinear terms of this system, we expand the SQUID cosine
potential to include non-quadratic corrections. Because
$\phi_{+}\ll 1$, it is enough to consider its forth-order terms in
the Lagrangian
\begin{equation}
\mathcal{L}=\frac{\Phi^{2}C_{s}}{(2\pi)^2}\dot{\phi}_{+}^{2}-\frac{E_{s}}{2}\phi_{+}^{2}+\frac{E_{s}}{24}\phi_{+}^{4}+\cdots.
\label{LSQUIDnon}
\end{equation}
The boundary condition in Eq.~(\ref{bound1}) now contains the
cubic term,
\begin{eqnarray}
&2&C_{s}\ddot{\psi}(d_{0},t)+\frac{\partial_{x}\psi(d_{0},t)}{L_{0}}\notag \\
&+&\frac{(2\pi)^2}{\Phi_{0}^{2}}E_{s}(\Phi_{\text{ext}})\left[\psi(d_{0},t)-\frac{1}{6}\psi^{3}(d_{0},t)\right]\!=0.
\label{cubic}
\end{eqnarray}
The cubic term not only relates the boundary equation with both
first and third-harmonic modes, but also produces a shift of the
resonant frequency, which depends on the photon number of the
resonator mode. Comparing Eq.~(\ref{cubic}) with
Eq.~(\ref{bound1}), we can roughly view the Josephson energy
$E_{1}(\Phi_{\text{ext}})$ to be slightly modified as
$$E_{1}(\Phi_{\text{ext}})\rightarrow
E_{1}(\Phi_{\text{ext}})[1-\frac{1}{6}\psi^{2}(d_{0},t)],$$ which
indicates that the nonlinear inductance
$L_{s1}(\Phi_{\text{ext}})$ now depends on the intracavity field
intensity $\psi^{2}(d_{0},t)$. Employing Eq.~(\ref{nonlinfre}),
and similar with the deviation in Ref.~\cite{Eichler2014}, the
quantized Hamiltonian of the fundamental mode with the self Kerr
nonlinearity can be approximately written as
\begin{equation}
H=\omega_{r0}a^{\dag}a+K_{D}a^{\dag}a^{\dag}aa, \label{directkerr}
\end{equation}
with the Kerr nonlinearity strength
\begin{eqnarray}
K_{D}&=&-\frac{\cos^{4}(\frac{\pi\omega_{r}}{2\omega_{0}})}{4 L_{s}
    (\Phi_{\text{ext}})}\left(\frac{2\pi\phi_{zpf}^{2}}{\Phi_{0}}\right)^{2} \notag \\
 &\approx&-
\frac{\pi e^{2}\omega_{r0}^{2}L_{t}}{8}\left[\frac{\pi
    L_{s}(\Phi_{\text{ext}})}{2L_{t}}\right]^{3}, \label{kerr}
\end{eqnarray}
where the quantized form of the field amplitude is
$$\psi_{1}(t)=\phi_{\text{zpf}}\left[a
\exp(-i\omega_{r0}t)+a^{\dagger}\exp(i\omega_{r0}t)\right],$$ with
$$\phi_{\text{zpf}}=\frac{2\pi}{\Phi_{0}}\sqrt{\frac{1}{2\omega_{r0}C_{t}}}$$
being the zero-point fluctuations of the flux
field~\cite{Wallquist06}. The \emph{self-Kerr nonlinearity} is due
to attaching the SQUID at the end the resonator. In our
discussion, the condition $L_{s}(\Phi_{\text{ext}})/L_{t}\ll 1$ is
always valid, and the \emph{Kerr strength} $K_{D}$ is proportional
to the cubic-order of the small parameter
$L_{s}(\Phi_{\text{ext}})/L_{t}$, which is much weaker than the
first-order effects [Eq.~(\ref{nonlinfre})]. In the following discussions, we consider this Kerr nonlinearity
effects on the qubit readout process.

We also need to check the validity of the quartic expansion approximation in Eq.~(\ref{LSQUIDnon}), which requires $\phi_{+}<1$. Considering the boundary condition in Eq.~(\ref{bound1}), one finds that the amplitude of the intracavity field at the position $d_{0}$ satisfies
\begin{equation}
\phi_{+}=\psi(d_{0},t)=2\phi_{\text{zpf}}\alpha_{c} \cos(kd_{0})<1,
\end{equation}
where $\alpha_{c}=\langle a\rangle$. Employing the relation $kd_{0}=(\pi\omega_{r})/(2\omega_{0})$ and according to the transcendental equation~(\ref{fstorder}), one finds the critical amplitudes given in Eq.~(\ref{Ncrt}).
In our discussions, we assume that $\omega_{r}$ is approximately around $\omega_{0}$, which results in a large critical photon number $n_{c}=|\alpha_{c}|^{2}$.

\subsection*{B.3 Resonator pure dephasing due to tunable boundary conditions}

Different from a frequency-fixed resonator, the mode frequency of
the SQUID-terminated resonator now depends on external parameters.
The bias noise of these control parameters leads to dephasing
processes of the resonator, which is similar to the qubit
case~\cite{Martinis03,Ithier05,Deppe07}. In our proposal, the mode
frequency is tuned via the flux bias through the SQUID and the
Josephson energy. The bias flux noise might come from the external
control lines, and the most important part is the $1/f$ noise.
Moreover, the noise in the critical current $I_{c}$ of each
junction may result in fluctuations of the Josephson energy via
the relation $E_{s0}=I_{c}\Phi_{0}/(2\pi)$~\cite{Koch071}.
Consequently, the resonator Hamiltonian can be formally written as
\begin{equation}
H_{r}=\omega_{r0}a^{\dagger}a+\left[\frac{\partial\omega_{r}}{\partial\Phi_{\text{ext}}}
\delta\Phi_{N}(t) +\frac{\partial\omega_{r}}{\partial I_{c}}
\delta\! I_{N}(t) \right]a^{\dagger}a, \label{Hrf}
\end{equation}
where $\delta\Phi_{N}(t)$ and $\delta\!I_{N}(t)$ are the flux and
critical current fluctuations around the static biases. For
convenience we set

$$f_{1}(t)=\frac{\partial\omega_{r}}{\partial\Phi_{\text{ext}}}
\delta\Phi_{N}(t), \qquad
f_{2}(t)=\frac{\partial\omega_{r}}{\partial I_{c}} \delta\!
I_{N}(t).$$ In the shifted frame of frequency $\omega_{r0}$,
decoherence processes can be defined via the time-dependent
off--diagonal operator
\begin{equation}
a(t)= \exp(-i\omega_{r0}t)\left\langle
\exp{\left[-i\int_{0}^{t}dt' \sum_{i=1,2}
    f_{i}(t')\right]}\right\rangle.
\end{equation}
The phase of the off-diagonal terms of $\langle a(t)\rangle$
acquires a random term $-i\int_{0}^{t}dt' \sum_{i=1,2} f_{i}(t')$.
The time average of the fluctuation correlation function is
defined by its noise power, which is expressed as
\begin{equation}
\langle
f_{i}(t)f_{i}(0)\rangle=\frac{1}{2\pi}\int_{-\infty}^{\infty}d\omega
S_{i}(\omega) \exp(i\omega t).
\end{equation}
Usually, the integrated noise is given by a Gaussian distribution,
and there is no correlation between these two different noise
sources. Similar to the discussions in
Refs.~\cite{Martinis03,Ithier05,Deppe07}, we obtain the following
relation
\begin{eqnarray}
&\left\langle \exp{\left[-i\int_{0}^{t}dt' \sum_{i=1,2}
    f_{i}(t')\right]}\right\rangle \notag\\
&=\exp{\left[-\frac{1}{2}\sum_{i=1,2}\left\langle\int_{0}^{t}dt' f_{i}(t') \int_{0}^{t}dt'' f_{i}(t'')\right\rangle\right]} \notag\\
&=\exp{\left[-\frac{1}{2}\int_{-\infty}^{\infty}
    \frac{d\omega}{2\pi}W(\omega)\sum_{i=1,2}S_{i}(\omega) \right]}.
\label{deco}
\end{eqnarray}
where $W(\omega)$ is the spectral weight function (see, e.g.,
Ref.~\cite{Martinis03}) given by
\begin{equation}
W(\omega)=\frac{\sin^{2}(\omega
    t/2)}{(\omega/2)^{2}}.
\end{equation}

The noise correlation function $S_{i}(\omega)$ determines the
decoherence behavior of the off-diagonal matrix elements. Given
that the correlation time of the noise is extremely short,
$S_{i}(\omega)$ is almost flat in the frequency domain, and the
corresponding line shape is Lorentzian with homogeneous
broadening~\cite{Koch071}. However, for the $1/f$ noise, its
correlation function is approximately described as $S_{i}(\omega)
\propto2\pi A_{i}^{2}/|\omega|$ with a singularity around
$\omega=0$~\cite{Martinis03}, where $A_{i}$ is the noise
amplitude. For simplification, we assume that the $1/f$ frequency
ranges of both flux and critical current noises are limited by the
same infrared ($\omega_{\text{ir}}$) and ultraviolet ($\omega_{\text{uv}}$)
cutoff. In this case, we solve the decoherence rate by
substituting $S_{i}(\omega)$ into Eq.~(\ref{deco}) to
obtain~\cite{Ithier05}:
\begin{eqnarray}
a(t)&\backsimeq&
\exp \Bigg\{-i\omega_{r0}t-\frac{1}{2}\bigg[\left(A_{1}\frac{\partial\omega_{r}}{\partial\Phi_{\text{ext}}}t\right)^{2} \notag \\
    &&+ A_{2}\left(\frac{\partial\omega_{r}}{\partial I_{c}}t
    \right)^{2}\bigg] |\ln(\omega_{\text{ir}}t)| \Bigg\}.
\label{eqf1}
\end{eqnarray}
We can roughly treat $\ln(\omega_{\text{ir}}t)$ as a constant, and find
that $\langle a(t)\rangle$ decays with time as $t^{2}$.
From Eq.~(\ref{eqf1}), the estimated dephasing rate
$\Gamma_{f,\Phi}$ and $\Gamma_{f,I}$ induced by the $1/f$ flux and
critical current noise are written as
\begin{equation}
\Gamma_{f,\Phi}=A_{1}\left(\frac{\partial\omega_{r}}{\partial\Phi_{\text{ext}}}\right),
\qquad \Gamma_{f,I}=A_{2}\left(\frac{\partial\omega_{r}}{\partial
    I_{c}}\right), \label{t2}
\end{equation}
respectively. In the following discussions, we evaluate the
decoherence effects induced by these bias noises.
\begin{center}
    \begin{figure*}[tbp]
        \centering
        \includegraphics[width=16.5cm]{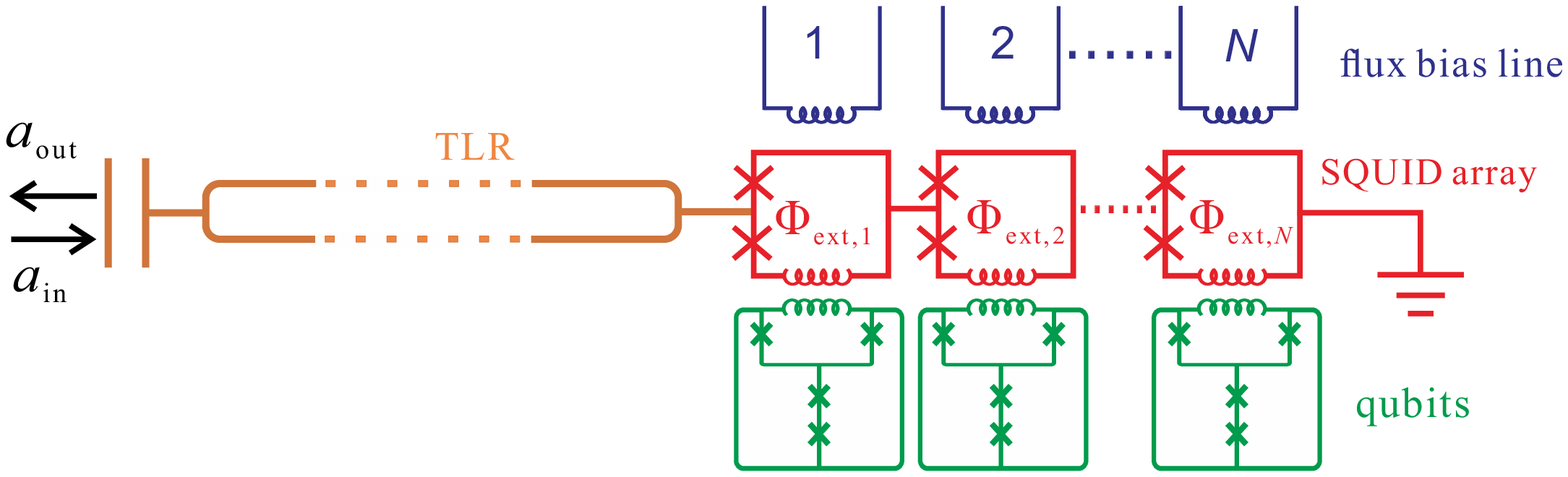}
        \caption{Schematics of a quarter-wavelength transmission-line resonator (TLR) terminated by an array of $N$ SQUIDs.
            The effective nonlinear inductance of the $j$th SQUID is controlled via the external flux
            $\Phi_{\text{ext},j}$, which contains a static part (produced by the $j$th bias line) and
            a perturbation part (produced by the circulating current of the $j$th flux qubit).  }
        \label{fig1s}
    \end{figure*}
\end{center}

\subsection*{B.4 Multi-SQUID terminated resonator}
As shown in Fig.~\ref{fig1s}, the one-dimensional transmission
line resonator (TLR) can also be terminated in its right side by a
series of $N$ dc SQUID~\cite{Sandberg2008,Eichler2014}. Each SQUID
can be tuned by via an independent external flux bias. The two
Josephson junctions of the $j$th SQUID are symmetric with
identical Josephson energy $E_{sj}$ and capacitance $C_{sj}$. The
effective Josephson energy is tuned with the external flux
$\Phi_{\text{ext},j}$ according to the relation
$E_{j}=2E_{sj}\cos(\pi \Phi_{\text{ext},j}/\Phi_{0})$, and the
nonlinear inductance is
$L_{sj}(\Phi_{\text{ext},j})=\Phi_{0}^{2}/[(2\pi)^2E_{j}]$. The
Lagrangian of the $j$th SQUID is

$$\mathcal{L}_{j}=\sum_{i=1,2}\left[\frac{\Phi_{0}^{2}}{2(2\pi)^{2}} C_{sj}(\dot{\phi_{ji}})^{2}-E_{sj}\cos{\phi_{ji}}\right],$$
where $\phi_{ji}$ is the phase difference of the $i$th junction in
the $j$th SQUID. Similar to the single SQUID case, we obtain the
boundary equation at the right-hand side of Eq.~(\ref{bound1})~\cite{Sandberg2008}:

\begin{subequations}
    \begin{gather}
    \psi(d_{0},t)=\sum_{j}^{N}\phi_{j+}(t),   \label{boundc1} \\
    2C_{s}\ddot{\psi}(d_{0},t)+\frac{(2\pi)^2}{\Phi_{0}^{2}}E_{j}(\Phi_{\text{ext},j})\phi_{j+}(t)+\frac{\partial_{x}
        \psi(d_{0},t)}{L_{0}}=0, \label{boundc2}
    \end{gather}
\end{subequations}
where $\phi_{j+}=(\phi_{j1}+\phi_{j2})/2$. By expanding the
left-hand side of Eq.~(\ref{boundc2}) with $\omega_r/\omega_{0}$
around $\pi/2$ to first order, we obtain
\begin{equation}
\omega_{r}=\omega_{0}\left[1-\frac{\sum_{j}^{N}L_{sj}(\Phi_{\text{ext},j})}{L_{t}}\right],
\label{1storder}
\end{equation}
from which we can find that the external flux
$\Phi_{\text{ext},j}$ through the $j$th SQUID determines its
nonlinear inductance independently. Their joint effect eventually
shifts the mode frequency to $\omega_{r}$. Similar to the
discussions of the single-SQUID case, we obtain the resonator
frequency $\omega_{r0}$ and the flux sensitivity $R_{j}$ of the
$j$th SQUID as

\begin{subequations}
    \begin{gather}
    \omega_{r0}=\omega_{0}\left[1-\sum_{j=1}^{N}\frac{L_{s}(\Phi_{\text{ext},j}^{0})}{L_{t}}\right],  \\
    R_{j}=\frac{\partial \omega_{r}}{\partial \Phi_{\text{ext},j}}\Big|_{\Phi_{\text{ext},j0}^{0}}
    =-\;\frac{\pi\omega_{0}L_{s}(\Phi_{\text{ext},j}^{0})}{\Phi_{0}L_{t}}\tan\!\left(\frac{\pi \Phi_{\text{ext},j}^{0}}{\Phi_{0}}\right).
    \end{gather}
    \label{Randfre1}
\end{subequations}

Note that the above discussions can also be applied to the
single-SQUID case by setting $N=1$. Assuming that the flux
perturbations of the $j$th SQUID are produced by the circulating
current of a single flux qubit as a quantum bus, then it is
possible to dispersively couple multiple qubits with a single
resonator. Employing this layout, we can achieve a \emph{multi-qubit QND
readout}.

\setcounter{equation}{0} \renewcommand{\theequation}{C\arabic{equation}}

\section{Circulating currents in the gradiometric flux qubit}

\begin{figure*}[tbp]
    \centering
    \includegraphics[width=16cm]{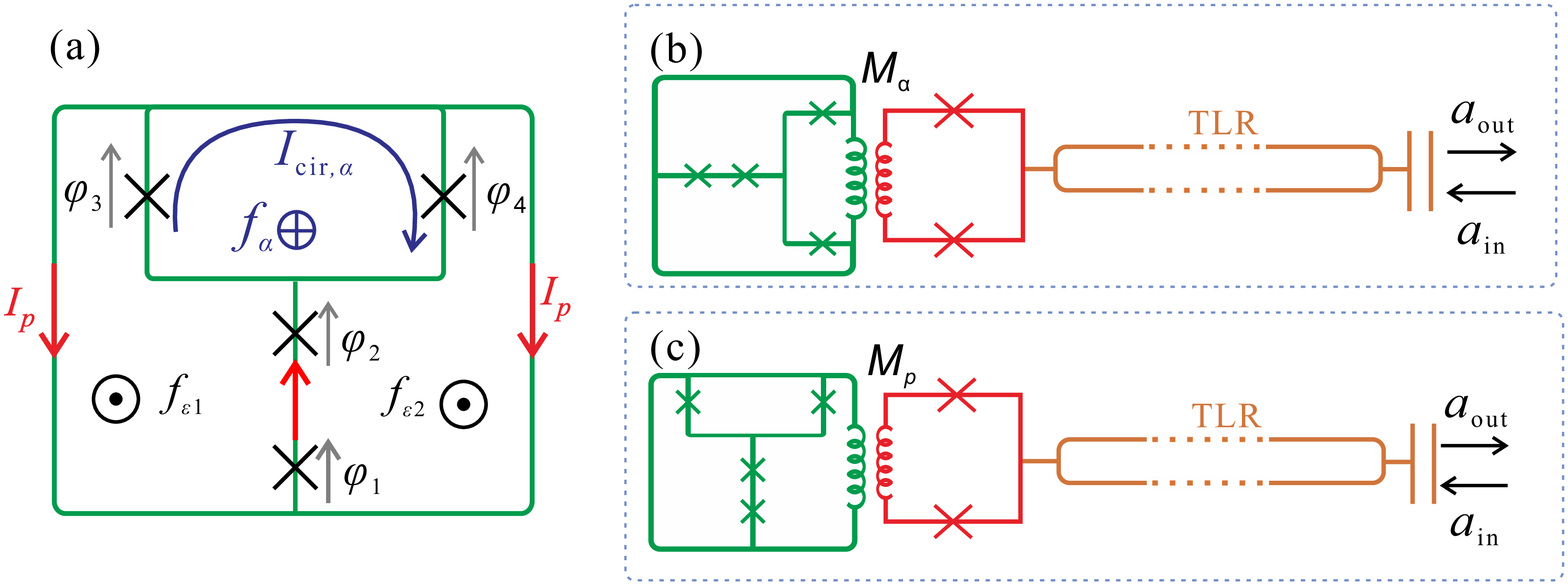}
    \caption{(a) Schematic diagram for a gradiometric flux qubit.
        The Josephson junctions are represented by the crosses.
        The phase difference across the $i$th junction is denoted by $\varphi_{i}$.
        The energy bias (energy gap) of the flux qubit is controlled via the reduced magnetic flux $f_{\epsilon1(2)}=\Phi_{\epsilon1(2)}/\Phi_{0}$ ($f_{\alpha}=\Phi_{\alpha}/\Phi_{0}$) through the two gradiometric loops
        ($\alpha$-loop). The persistent current $I_{p}$ (red arrows)
        is split into two symmetrical parts, which circulate
        around two identical gradiometric loops. Another
        quantized super-current $I_{\text{cir},\alpha}$ circulates in the $\alpha$-loop, which
        is usually employed to create the longitudinal coupling.
        (b,c) Schematics of a flux qubit interacting with a
        SQUID-terminated resonator via mutual inductance $M$ and
        circulating current $I$: (b) $M_{\alpha}$ and
        $I_{\text{cir},\alpha}$, (c) $M_{p}$ and $I_{p}$.}
    \label{fig2s}
\end{figure*}

As shown in Fig.~\ref{fig2s}(a), the gap-tunable flux qubit has a
gradiometric topology by adopting an eight-shaped design, and the
small $\alpha$-junction is replaced by a SQUID (the
$\alpha$-loop). The gradiometric structure splits the persistent
current symmetrically. This special geometric arrangement allows
one to control the gap value $\alpha$ via the external flux
$f_{\alpha}$ without disturbing the energy
bias~\cite{Paauw09,Fedorov10,Schwarz2013}. We assume that the two
junctions (with a gauge-invariant phase difference
$\varphi_{1,2}$) in the main loop have the same Josephson energy
$E_{J}$ and capacitance $C$. The other two junctions in the SQUID
loop (with a gauge-invariant phase difference $\varphi_{3,4}$),
are also identical but with smaller Josephson energies and
capacitances by a factor $\alpha_{0}$ compared to the junctions in
the main loop. Because the loop inductance is usually much smaller
than the effective nonlinear junction, we neglect the phase
accumulated along each loop circumference. Therefore, the fluxoid
quantization conditions of the $f_{\alpha}$, $f_{\epsilon1}$ and
$f_{\epsilon2}$ loop are~\cite{Paauwthesis,Schwarz15doc}:
\begin{subequations}
    \begin{align}
    \varphi_{3}-\varphi_{4}+2\pi f_{\alpha}=2\pi N_{\alpha},     \\
    \varphi_{3}+\varphi_{1}+\varphi_{2}+2\pi f_{\epsilon1}=2\pi N_{\epsilon1},    \\
    -\varphi_{4}-\varphi_{1}-\varphi_{2}+2\pi f_{\epsilon2}=2\pi
    N_{\epsilon2},
    \end{align}
\end{subequations}
where $N_{\alpha}$ and $N_{\epsilon1,\epsilon2}$ are the integer
numbers of the trapped fluxoids, and $f_{\alpha,\epsilon1(2)}=
\Phi_{\alpha,\epsilon1(2)}/\Phi_{0}$ with $\Phi_{\alpha}$
($\Phi_{\epsilon1(2)}$) being the external flux through the
$f_{\alpha}$ ($f_{\epsilon1,2}$) loop. We assume $N_{\alpha}=0$.
By setting $n=N_{\epsilon2}-N_{\epsilon1}$ and
$f_{\epsilon}=\Phi_{\epsilon}/\Phi_{0}=f_{\epsilon1}-f_{\epsilon2}$,
the above boundary conditions reduce the freedom of the systems
and, thus, $\varphi_{3,4}$ can be given in terms of
$\varphi_{1,2}$
\begin{equation}
\varphi_{3,4}=-\pi(n+ f_{\epsilon})-(\varphi_{1}+\varphi_{2})\mp\pi f_{\alpha}.
\label{phi34}
\end{equation}
The Josephson energy (or the potential energy) for this
four-junction system as a function of $\varphi_{1}$ and
$\varphi_{2}$ is expressed as:
\begin{eqnarray}
U/E_{J}&=&2+2\alpha_{0}-\cos{\varphi_{1}}-\cos{\varphi_{2}} \notag \\
&-&2\alpha_{0} \cos(\pi
f_{\alpha})\cos{[\varphi_{1}+\varphi_{2}+\pi(n+f_{\epsilon})]}.
\end{eqnarray}
We now consider the charging energy stored in the capacitances of
the four junctions in this circuit, and the kinetic energy has the
form
\begin{eqnarray}
T&=&\frac{1}{2}\left(\frac{\Phi_{0}}{2\pi}\right)^{2}
\sum_{i=1}^{4}C_{i}\dot{\varphi}_{i}^{2} \notag \\
&=&\left(\frac{\Phi_{0}}{2\pi}\right)^{2}\!C\!\left[\frac{(1+2\alpha_{0})}{4}(\dot{\varphi}_{1}^{2}
+\dot{\varphi}_{2}^{2})+\alpha_{0}\dot{\varphi}_{1}\dot{\varphi}_{2}\right],
\label{kinetic}
\end{eqnarray}
where we have employed the relation
$\dot{\varphi}_{3,4}=-(\dot{\varphi}_{1}+\dot{\varphi}_{2})$. The
Lagrangian for the whole circuit is $\mathcal{L}=T-U$, from which
we obtain the canonical momentum $p_{i}=\partial
\mathcal{L}/(\partial \dot{\varphi}_{i})$ as the conjugate to the
coordinate $\varphi_{i}$. Therefore, by employing a Legendre
transformation, the corresponding Hamiltonian is written as
\begin{eqnarray}
H&=&\sum_{i=1,2}p_{i}\dot{\varphi}_{i}-\mathcal{L}  \notag\\
&=&\frac{4E_{c}}{1+4\alpha_{0}}[(1+2\alpha_{0})(p_{1}^{2}+p_{2}^{2})-4\alpha_{0} p_{1}p_{2})]  \notag\\
&&+E_{J}\big\{2+2\alpha_{0}-\cos{\varphi_{1}}-\cos{\varphi_{2}} \notag \\
&&-2\alpha_{0} \cos(\pi
f_{\alpha})\cos{[\varphi_{1}+\varphi_{2}+\pi(n+f_{\epsilon})]}\big\}.
\label{Ham1}
\end{eqnarray}
To quantize the above Hamiltonian, we introduce the commutation
relation $[\varphi_{i},p_{j}]=i \delta_{ij}$ with $p_{i}=-i
\partial/\partial \varphi_{i}$.

For a flux qubit, to minimize the dephasing induced by the flux
noise in the main-loop, one usually operates the flux qubit at its
degeneracy point with $f_{\epsilon}=0$. Moreover, under the
condition $0<2\alpha_{0} \cos(\pi f_{\alpha})<1$, the potential
$U$ has a double-well shape. The eigenproblem described by
Eq.~(\ref{Ham1}) can be numerically solved in the plane-wave
basis~\cite{Orlando1999}. As discussed in Ref.~\cite{Paauwthesis},
the two lowest energy-levels are well-separated from all the
higher ones. The ground state $|g\rangle$ and the first excited
state $|e\rangle$ are, respectively, symmetric and anti-symmetric
along the axis $\varphi_{+}=\varphi_{1}+\varphi_{2}$, and can be
approximately expressed as~\cite{Schwarz15doc}:

\begin{subequations}
    \begin{align}
    |g\rangle=\frac{1}{\sqrt{2}}(|+I_{p}\rangle+|-I_{p}\rangle),   \\
    |e\rangle=\frac{1}{\sqrt{2}}(|+I_{p}\rangle-|-I_{p}\rangle),
    \end{align}
\end{subequations}
where $|\pm I_{p}\rangle$ are the two persistent-current states of
the opposite directions in the
main-loop~\cite{Orlando1999,Mooij99}. To calculate the circulating
current in the $\alpha$-loop, we focus our attention on the
supercurrent through the junctions 3 and 4, which is expressed as
\begin{equation}
I_{3,4}=\frac{2\pi \alpha_{0} E_{J}}{\Phi_{0}}
\sin(\varphi_{3,4})=I_{c}\sin(\varphi_{3,4}). \label{I34}
\end{equation}
Employing the relation in Eq.~(\ref{phi34}) and expanding
$I_{3,4}$ in the basis of $|g\rangle$ and $|e\rangle$, we obtain
the current operator for the junctions 3 and 4 as follows
\begin{eqnarray}
&&I_{3(4)}=\left(
\begin{matrix}
I_{3(4),ee} & I_{3(4),eg}\\
I_{3(4),ge} & I_{3(4),gg}\\
\end{matrix}
\right) \notag \\
&&\!=I_{c}\!\left(
\begin{matrix}
\langle e|\sin(\varphi_{+}\!\pm\!\pi f_{\alpha})|e\rangle &\langle e|\sin(\varphi_{+}\!\pm\!\pi f_{\alpha})|g\rangle\\
\langle g|\sin(\varphi_{+}\!\pm\!\pi f_{\alpha})|e\rangle &\langle g|\sin(\varphi_{+}\!\pm\!\pi f_{\alpha})|g\rangle\\
\end{matrix}
\right)\!,
\end{eqnarray}
Because $|g\rangle$ ($|e\rangle$) is of even (odd) parity with
respect to $\varphi_{+}$ at the degeneracy point, it is easy to
verify that
\begin{equation}
    I_{3,ii}=-I_{4,ii}=\sin(\pi f_{\alpha})I_{c} \langle i|\cos(\varphi_{+})|i\rangle
\end{equation}
for $i=e,\,g$.
Therefore, given that the qubit is in its excited (ground) state,
the average current of the junctions 3 and 4 are of opposite
signs, and they generate a circulating current with amplitude
$I_{3,ee}$ ($I_{3,gg}$) in the $\alpha$-loop. For the off-diagonal
terms, it can be easily verified that
\begin{equation}
I_{3,eg}=I_{4,eg}=\cos(\pi f_{\alpha})I_{c} \langle
e|\sin(\varphi_{+})|g\rangle, \label{I3eg}
\end{equation}
which is, in fact, equal to the persistent-current $I_{p}$ of the
main loop (related to the $\sigma_{x}$ operator). Therefore, the
circulating current operator in the $\alpha$-loop and persistent
current operator in the main-loop is expressed as

\begin{subequations}
    \begin{gather}
    I_{\text{cir},\alpha}=\frac{I_{3}-I_{4}}{2}=I_{+}I_{0}+I_{-}\sigma_{z},   \\
    I_{p}'=I_{3}+I_{4}=2I_{3,eg}\left(|e\rangle\langle
    g|+|g\rangle\langle e|\right)=I_{p}'\sigma_{x}, \label{ICR}
    \end{gather}
\end{subequations}
where $$I_{+}=(I_{3,ee}+I_{3,gg})/2, \quad I_{-}=(I_{3,ee}-I_{3,gg})/2,$$ and $I_{0}$ is the identity
operator. The standard definition of the SQUID circulating current
in Refs.~\cite{Paauwthesis,Wang11} also gives the same form for
$I_{\text{cir},\alpha}$. From Eq.~(\ref{Ham1}) we find that the
flux qubit is controlled by the external fluxes $f_{\alpha}$ and
$f_{\epsilon}$. Assuming that $n=1$ and the flux qubit is
prebiased at the optimal point
$\{f_{\alpha},f_{\epsilon}\}=\{f_{\alpha0},0\}$, the circulating
currents $I_{p}'$ and $I_{\text{cir},\alpha}$ can also be derived
via a thermodynamic relation~\cite{Orlando1999,Paauwthesis}.
Specifically, by considering the flux perturbations $\delta\!
f_{\alpha}$ and $\delta\! f_{\epsilon}$, we can rewrite the
Hamiltonian in Eq.~(\ref{Ham1}) as
\begin{equation}
H(f_{\alpha},f_{\epsilon})=H(f_{\alpha0},f_{\epsilon0})+\frac{\partial
    H}{\partial f_{\alpha}}\delta \! f_{\alpha}+ \frac{\partial
    H}{\partial f_{\epsilon}}\delta \! f_{\epsilon}. \label{partialH}
\end{equation}
In the basis of $|e\rangle$ and $|g\rangle$, it can be easily
verified that the following thermodynamic relations
hold~\cite{Orlando1999,Paauwthesis}
\begin{equation}
\frac{\partial H}{\partial
    f_{\alpha}}=I_{\text{cir},\alpha}\Phi_{0};\qquad \frac{\partial
    H}{\partial f_{\epsilon}}=\frac{I_{p}'}{2}\Phi_{0}.
\label{thermalre}
\end{equation}
Therefore, we can rewrite Eq.~(\ref{partialH}) as
\begin{eqnarray}
H=\frac{1}{2} \omega_{q}
\sigma_{z}+\frac{I_{p}'}{2}\sigma_{x}\delta\Phi_{\epsilon}
+(I_{+}I_{0}+I_{-}\sigma_{z})\delta\Phi_{\alpha}. \label{Hpf}
\end{eqnarray}
where $\delta\Phi_{\alpha}=\delta\! f_{\alpha}\Phi_{0}$ and
$\delta\Phi_{\epsilon}=\delta\! f_{\epsilon}\Phi_{0}$. The
circulating currents $I_{\text{cir},\alpha}$ and $I'_{p}$, obtained
from the thermodynamic relation Eq.~(\ref{thermalre}) and the
definitions in Eq.~(\ref{ICR}) lead to the same results. For the
flux qubit working at its degeneracy point, the qubit transition
frequency $\omega_{q}$ between $|e\rangle$ and $|g\rangle$ is
determined by the control flux $f_{\alpha}$, and the dephasing
resulting from the flux noise in $f_{\epsilon}$ vanishes to
first-order. The effective persistent-current circulating in each
symmetric main loop is divided by two due to the gradiometric
topology, i.e., $I_{p}=I_{p}'/2$.

In this gap-tunable flux qubit, the persistent current $I_{p}$ of
the flux qubit is widely employed to create ($\sigma_{x}-$type)
dipole couplings. The circulating current difference $I_{-}$ in
the $\alpha$-loop, can be employed to induce longitudinal coupling
($\sigma_{z}-$type)~\cite{Schwarz2013}. As shown in
Fig.~\ref{fig2s}(a) and \ref{fig2s}(b), the circulating currents
$I_{\text{cir},\alpha}$ and $I_{p}$ can produce a flux
perturbation through the SQUID of the resonator via the mutual
inductances $M_{\alpha}$ and $M_{p}$, respectively, which changes
the effective length of the resonator. Specifically, in the basis
of $|e\rangle$ and $|g\rangle$, the interaction between
$I_{\text{cir},\alpha}$ and the SQUID-terminated resonator
corresponds to NPDC.

Note that both circulating currents $I_{\text{cir},\alpha}$ and
$I_{p}$ naturally enhance the qubit sensitivity to flux noises.
The $1/f$-type flux noise of the $\alpha$-loop leads to the
broadening of the qubit transition frequency $\omega_{q}$, which
corresponds to a pure dephasing process ($T_{2}$). The flux noise
through two gradiometric loops affects the qubit via the
persistent-current operator $I_{p}\sigma_{x}$, which results in
the energy-relaxation process ($T_{1}$). Similar to discussion in
Ref.~\cite{Martinis03,Ithier05,Koch071}, the relaxation and
dephasing rates can be approximately written as
\begin{eqnarray}
\Gamma_{1}&=&\frac{1}{T_{1}}\backsimeq \left( \frac{\partial
    H}{\partial f_{\epsilon}}\right)^{2}S_{\bot}(\omega_{q})=
\left(I'_{p}\Phi_{0} \right)^{2}S_{\bot}(\omega_{q}), \notag \\
\Gamma_{f}&=&\frac{1}{T_{2}}\backsimeq \left(\frac{\partial
    H}{\partial f_{\alpha}}\right)A_{\alpha}=
A_{\alpha}\left(I_{-}\Phi_{0}\right), \label{qubitde}
\end{eqnarray}
where  $S_{\bot}(\omega_{q})$ is the noise power at the qubit
frequency, and $A_{\alpha}$ is the amplitude of the $1/f$-type
flux noise in the $\alpha$-loop. Note that the nonzero current
difference $I_{-}$ makes the qubit sensitive to the $1/f$ noise in
the $\alpha$-loop. As in the following discussion, for the flux
qubit, the amplitude of $I_{-}$ is usually lower than the
persistent current $I_{p}$ by about one order of magnitude. Moreover, the
experimental results reported in Ref.~\cite{Fedorov10} indicate
that the flux noise  $A_{\alpha}$ might be much smaller than that
of the main loop. Therefore the dephasing rate induced by $I_{-}$
is possibly much slower than that in the case when the qubit is
operating far away from its optimal point.

\setcounter{equation}{0} \renewcommand{\theequation}{D\arabic{equation}}

\section{Numerical results on coupling strength, nonlinearity, and decoherence}

We now discuss a set of possible parameters for the
SQUID-terminated nonlinear resonator. Our discussions are mainly
based on the experimental parameters in
Refs.~\cite{Pogorzalek17,Eichler18}. We first consider a
$\lambda/4$ resonator with fixed frequency
$\omega_{0}\simeq2\pi\times6~\text{GHz}$ and total inductance
$L_{t}=10~\text{nH}$. A rapid photon escape rate $\kappa$ enhances
the speed of the qubit readout, and we set $\kappa/(2\pi)\simeq
16~\text{MHz}$ in the following discussion. By assuming
$E_{s0}=2\pi\times2.5~\text{THz}$, the flux sensitivity and the
self-Kerr nonlinearity strength changing with the control flux
have been shown in Fig.~\ref{fig2m}. The flux
sensitivity is about
$R/(2\pi)\simeq16~\text{MHz}/(\text{m}\Phi_{0})$ with the Kerr
nonlinearity $K_{D}/(2\pi)\simeq110~\text{kHz}$ at
$\Phi_{\text{ext}}\simeq 0.48\Phi_{0}$.

The flux (critical current) noise amplitude of the SQUID attached
to the resonator can be set as $A_{1}=5\times10^{-6}\Phi_{0}$
($A_{2}=5\times10^{-6}I_{c}$)~\cite{Koch071}. Employing these
parameters, the dephasing rates in Eq.~(\ref{t2}) are calculated
as $\Gamma_{f,\Phi}/(2\pi)\simeq75~\text{kHz}$ and
$\Gamma_{f,I}/(2\pi)\simeq30~\text{kHz}$. Therefore the reduction
of $T_{2}$ due to these two dephasing processes can be neglected
when compared with $\kappa$ in our discussions.
\begin{figure*}[tbp]
    \centering
    \includegraphics[width=18cm]{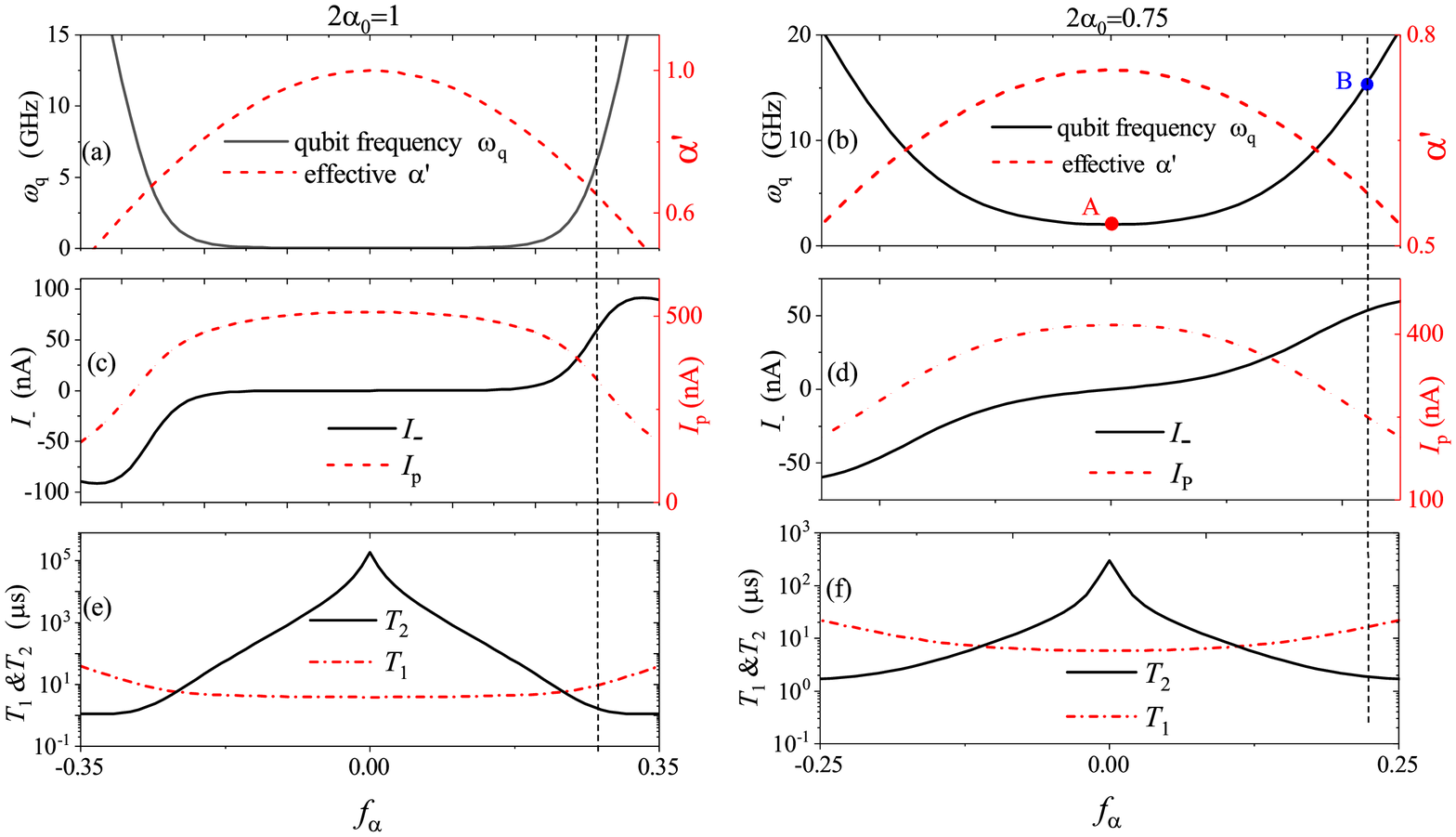}
    \caption{
        Flux-qubit parameters versus the external control flux
        $f_{\alpha}$ for $2\alpha_{0}=1$ (left panels) and
        $2\alpha_{0}=0.75$ (right panels). These parameters include: (a,b)
        the effective gap value $\alpha'$, given in
        Eq.~(\ref{alphaPrime}), (red dashed curve) and the qubit frequency
        $\omega_{q}$ (black solid curve); (c,d) Circulating-current
        amplitudes $I_{-}$ (red dashed curve) and $I_{-}$ (black solid
        curve); and (e,f) the energy-relaxation times $T_{1}$ (red dash-dotted
        curve) and decoherence $T_{2}$ (black curve). In
        the left panels, the vertical line is at position
        $f_{\alpha}\simeq0.27$. In (b), the points A and B correspond to
        $f_{\alpha}\simeq0$ and $f_{\alpha}\simeq0.22$, respectively. Here
        we assume $E_{J}/(2\pi)=320~\text{GHz}$ and $E_{J}/E_{C}=70$. The
        flux-noise amplitudes are set as
        $S_{\bot}(\omega_{q})=5\times10^{-10}~\text{s}$ and
        $A_{\alpha}=5\times10^{-6}$ according to
        Refs.~\cite{Ithier05,Koch071}.}
    \label{fig3s}
\end{figure*}

To view the whole circuit in Fig.~\ref{fig2s}(a) as a flux qubit,
the effective gap value
\begin{equation}
\alpha'=2\alpha_{0} \cos(\pi f_{\alpha})
\label{alphaPrime}
\end{equation}
should be in the range $0.5<\alpha'<1$~\cite{Orlando1999}, so that
the double-well potential approximation is valid. In
Fig.~\ref{fig3s}, we plot the qubit parameters changing with the
external control flux $f_{\alpha}$ by setting $2\alpha_{0}=1$
(left panel) and $2\alpha_{0}=0.75$ (right panel), respectively.
As shown in Fig.~\ref{fig3s}(a) and (b), the qubit frequency
$\omega_{q}$ can be tuned in a wide range, when $f_{\alpha}$ is
biased to be nonzero. The slope of $\omega_{q}$ changing with
$f_{\alpha}$ is proportional to the circulating-current difference
$I_{-}$ in the $\alpha$-loop [Eq.~(\ref{thermalre})].

As shown in Fig.~\ref{fig3s}(a), the two-energy-level structure
vanishes (i.e., $\omega_{q}=0$) at $f_{\alpha}=0$ for
$2\alpha_{0}=1$. To obtain a qubit energy-level structure, we need
to bias $f_{\alpha}$ far away from zero, for example, to the
dashed-line position, where there is an effectively nonzero
circulating current $I_{-}\simeq 60~\text{nA}$
[Fig.~\ref{fig3s}(c)]. When we keep on biasing $f_{\alpha}$,
$I_{-}$ becomes larger. As described by Eq.~(\ref{qubitde}),
increasing $I_{-}$ leads to a reduction of the pure dephasing time
$T_{2}$. In Fig.~\ref{fig3s}(e), we find that the pure dephasing
time $T_{2}$ decreases to about $1~\mu\text{s}$ at the dashed line
position. In a gate operation, one may need a much longer qubit
dephasing time. The experimental results in Ref.~\cite{Fedorov10}
indicate that the flux noise in the $\alpha$-loop has a much lower
amplitude than that in the main loop, and it is possible to obtain
longer $T_{2}$ in experiments by reducing the $(1/f)$ noise
amplitude $A_{\alpha}$.

Here we discuss another approach to increase $T_{2}$. In fact, by
setting $2\alpha_{0}<1$, the qubit is insensitive to the first
order of the flux noise in the $\alpha$-loop at $f_{\alpha}=0$,
and the examples with $2\alpha_{0}=0.75$ are plotted in the right
panel of Fig.~\ref{fig3s}. At $f_{\alpha}=0$ (point A), the qubit
frequency is $\omega_{q}/(2\pi)\backsimeq2~\text{GHz}$ with
$I_{-}=0$. Because $I_{-}=0$, the qubit is insensitive, to first
order of the $1/f$ flux noise, and $T_{2}$ is much longer than
$1~\mu\text{s}$. When employing this qubit for quantum-information
processing, one can operate it at the point A with much longer
dephasing time. Once the qubit state is to be measured, the flux
$f_{\alpha}$ is adiabatically biased away from zero without
damping a given qubit state. As shown in Fig.~\ref{fig3s}(d), the
circulating current $I_{\text{cir},\alpha}$ increases with
$|f_{\alpha}|$. At $f_{\alpha}\backsimeq0.22$ (Point B),
$I_{\text{cir},\alpha}\backsimeq50~\text{nA}$ and the dephasing
time is about $T_{2}\sim1~\mu\text{s}$. As discussed in
Ref.~\cite{Jeffrey14}, the qubit-readout time can be finished in
tens of $\text{ns}$ and therefore it is possible to perform
several measurements within $T_{2}$. After finishing the
measurements, one can adiabatically reset the flux bias
$f_{\alpha}=0$ with a longer dephasing time for further quantum
information processing. When considering the readout via
changing the qubit frequency over such a large range, we should
reconsider the parameters of the flux qubit carefully, e.g., the
transition effects to higher-energy levels, and the degradation of
the relaxation time $T_{1}$ and coherence time $T_{2}$. The
breakdown of the adiabatic approximation, which indicates the
coherence loss of the flux qubit, leads to a significantly lower
readout fidelity.

As shown in Fig.~\ref{fig2s}(b), assuming that the qubit interacts
with the resonator via mutual inductance $M_{\alpha}$, the
circulating current $I_{\text{cir},\alpha}$ produces a small
deviation part $\delta \Phi_{\text{ext}}= M_{\alpha}
I_{\text{cir},\alpha}$, which can be detected by the resonator
with a flux sensitivity $R$. Thus, the flux qubit can be coupled
to the SQUID-terminated resonator. To enhance their coupling
strength, we should employ a large mutual inductance to sense the
circulating current. Assuming the mutual inductance between the
$\alpha$-loop and the SQUID of the resonator is $M_{\alpha}$, the
Hamiltonian for the whole system can be written as
\begin{eqnarray}
H_{D}&=&\frac{\omega}{2}
\sigma_{z}+\omega_{r0}a^{\dag}a \notag \\
&+&RM_{\alpha}(I_{+}I_{0}
+I_{-}\sigma_{z})a^{\dag}a+K_{D}a^{\dag}a^{\dag}aa     \notag  \\
&=&\frac{\omega}{2} \sigma_{z}+
\omega_{r}'a^{\dag}a+\chi_{z}^{D}\sigma_{z}a^{\dag}a+K_{D}a^{\dag}a^{\dag}aa,
\label{halphadis}
\end{eqnarray}
where $\chi_{z}^{D}=RM_{\alpha}I_{-}$ is the NPDC strength, and
$\omega_{r}^{'}=\omega_{r0}+RM_{\alpha}I_{+}$ is the renormalized
mode frequency. One can find that this coupling has no relation to
the dipole-field interactions. This qubit readout based on the
Hamiltonian~(\ref{halphadis}) can be denoted as \emph{ideal QND}
\emph{measurement} because $H_{D}$ commutes with the qubit
operator $\sigma_{z}$.

As depicted in Fig.~\ref{fig3s}, we set
$I_{\text{cir},\alpha}\backsimeq60~\text{nA}$ and
$I_{p}\backsimeq300~\text{nA}$ in our discussion. To obtain strong
coupling strengths, we can employ the kinetic inductance by
sharing a qubit loop branch with the resonator SQUID. The kinetic
mutual inductance is about $1\sim3~\text{pH}/\mu\text{m}$, and can
still be enhanced by reducing the wires cross-section
area~\cite{Meservey1969,Paauwthesis,Schwarz15doc}. The mutual
inductance is about $15~\text{pH}$ with a shared loop length
$\sim5~\mu\text{m}$. Employing these parameters, we find that the
coupling strengths are $\chi_{z}^{D}\backsimeq7~\text{MHz}$ and
$\chi_{x}^{D}\backsimeq35~\text{MHz}$, respectively.

In the readout experiment with the IDC in Ref.~\cite{Jeffrey14},
the Jaynes-Cummings coupling strength is about
$g_{x}/(2\pi)=90~\text{MHz}$ with detuning $\Delta_{d}\simeq
1~\text{GHz}$, and the calculated IDC strength is about
$\chi_{z}^{I}/(2\pi)\simeq8~\text{MHz}$ with the qubit-state
dependent Kerr nonlinearity $K_{I}/(2\pi)\simeq 65~\text{kHz}$. We
find that it is reasonable to assume that
$\chi_{z}^{I}=\chi_{z}^{D}$ and $K_{D}\simeq K_{I}$ in our
discussions.

Moreover, we have plotted the energy relaxation time $T_{1}$
changing with $f_{\alpha}$. It can be found that $T_{1}$ varies
over a much smaller scale that $T_{2}$. By assuming the noise
power spectrum at the qubit frequency
$S_{\bot}(\omega_{q})=(5\times10^{-10})^{2}~\text{s}$~\cite{Ithier05},
the relaxation time is around
$\Gamma_{1}^{-1}\backsimeq9~\mu\text{s}$, which is of the same
order as the experimental results~\cite{Stern14}. In a qubit
readout proposal based on IDC, the resonator usually has a
quick decay rate. By setting the photon escaping rate
$\kappa/(2\pi)=16~\text{MHz}$ and $\lambda=0.1$, 
the energy relaxation time due to the Purcell effect $T_{p}=\Gamma_{p}^{-1}$ 
is approximately $1~\mu\text{s}$. Because
$\Gamma_{p}\gg \Gamma_{1}$, and it is reasonable to assume that
the qubit decay is mainly limited by Purcell effects.

\setcounter{equation}{0} \renewcommand{\theequation}{E\arabic{equation}}

\section{Dispersive qubit readout without Purcell decay}

From the discussions above, we find that the Kerr nonlinearity is
involved in a qubit readout for both the IDC and NPDC readouts.
However, these two nonlinearities are due to two different
mechanisms: $K_{D}$ is due to attaching a nonlinear SQUID in the
measurement resonator, while $K_{I}$ results from qubit dressing
effects via the dipole-field coupling.

As shown in Fig.~\ref{fig1m}, at $t=0$ we apply an
incident field $a_{\text{in}}$ in the left port at the shifted
frequency of the resonator. In the interaction picture, the
Langevin equations of the resonator operator, governed by
Eq.~(\ref{inducedis}) (the IDC readout) and Eq.~(\ref{halphadis})
(the NPDC readout) can, respectively, be written as
\begin{eqnarray}
\frac{da(t)}{dt}&=&-i\chi_{z}^{I}\sigma_{z} a(t)-2iK_{I}\langle
n(t)\rangle \sigma_{z} a(t) \notag \\
&-&\frac{1}{2}\kappa a(t)
-\sqrt{\kappa}a_{\text{in}}(t),
\label{inducedLan}\\
\frac{da(t)}{dt}&=&-i\chi_{z}^{D}\sigma_{z} a(t)-2iK_{D}\langle
n(t)\rangle a(t)  \notag \\
&-&\frac{1}{2}\kappa a(t)
-\sqrt{\kappa}a_{\text{in}}(t), \label{directLan}
\label{directnn}
\end{eqnarray}
where $\langle n(t)\rangle=\langle a^{\dag}(t)a(t)\rangle$ is the
time-dependent photon number in the resonator. The \emph{Kerr term}
$K_{I}$ in Eq.~(\ref{inducedLan}) is dependent on the qubit state,
i.e., is related to the Pauli operator $\sigma_{z}$; while the
\emph{Kerr nonlinearity} $K_{D}$ in Eq.~(\ref{directLan}) is a standard
Kerr term. This input field
$a_{\text{in}}(t)=\alpha_{\text{in}}-d_{\text{in}}(t)$  is assumed
to be characterized by its mean value (a coherent drive)
$\alpha_{\text{in}}=-\epsilon \exp(i\theta_{d})/\sqrt{\kappa}$ and
a fluctuation part $d_{\text{in}}(t)$. To compare the qubit
readout process for these two different mechanisms, we assume
$\chi_{z}^{I}=\chi_{z}^{D}=\chi_{z}$ and $K_{D}=K_{I}=K$ in the
following discussion.

Below we start from the ideal readout without the Kerr
nonlinearity, i.e., $K=0$, and give an analytical form for the
measurement fidelity. After that, we reconsider the nonlinear
effects in these two cases.

\subsection*{E.1 Ideal readout: Measurement without Kerr nonlinearity}

By setting $K_{D}=K_{I}=K=0$, we obtain the same linear Langevin
differential equation from both Eqs.~(\ref{inducedLan}) and
(\ref{directLan}). The average part of the output field is
obtained from the input-output boundary condition
$$\alpha_{\text{out}}=\sqrt{\kappa} \alpha_{r}(t)-\epsilon
\exp(i\theta_{d})/\sqrt{\kappa},$$ where $\alpha_{r}(t)$ is the
average field of the resonator, and is derived by formally
integrating the Langevin differential equation~\cite{Didier152}:
\begin{eqnarray}
\alpha_{r}(t)&=&\frac{\epsilon \sqrt{\kappa} \exp\left[i(\theta_{d}-\langle
    \sigma_{z}\rangle
    \theta_{q})\right]}{\sqrt{\frac{1}{4}\kappa^{2}+(\chi_{z}\langle
        \sigma_{z}\rangle)^{2}}}    \notag \\
&&\times\left\{1-\exp\left[-(i\chi_{z} \langle \sigma_{z}\rangle+\frac{1}{2}\kappa)t\right]\right\},
\label{alphabar}
\end{eqnarray}
where $\theta_{q}=\arctan({2\chi_{z}/\kappa})$ is the rotating
angle of the output field due to the dispersive coupling.
The average intracavity photon number is written as in Eq.~(\ref{intraphoton}).
The output fluctuation part $d_{\text{out}}(\omega)$ in
Fourier space can also be obtained from the Langevin differential
equation, and is expressed as
\begin{equation}
d_{\text{out}}(\omega)=\left[1-\frac{\kappa}{i(\omega+\chi_{z}\langle
    \sigma_{z}\rangle)+\frac{1}{2}\kappa}\right]d_{\text{in}}(\omega)
\label{fluctuation}
\end{equation}
One find that Eq.~(\ref{fluctuation}) leads to completely
different expressions for different types of input noise
$d_{\text{in}}(\omega)$, (e.g., the vacuum, single-, and
multi-mode squeezed vacuum). For simplicity, we assume that
$d_{\text{in}}(t)$ is the vacuum without squeezing, and satisfies
the correlation relation $\langle d_{\text{in}}(\omega)
d_{\text{in}}^{\dag}(\omega')\rangle=\delta (\omega+\omega')$.
\begin{figure*}[tbp]
    \centering
    \includegraphics[width=18cm]{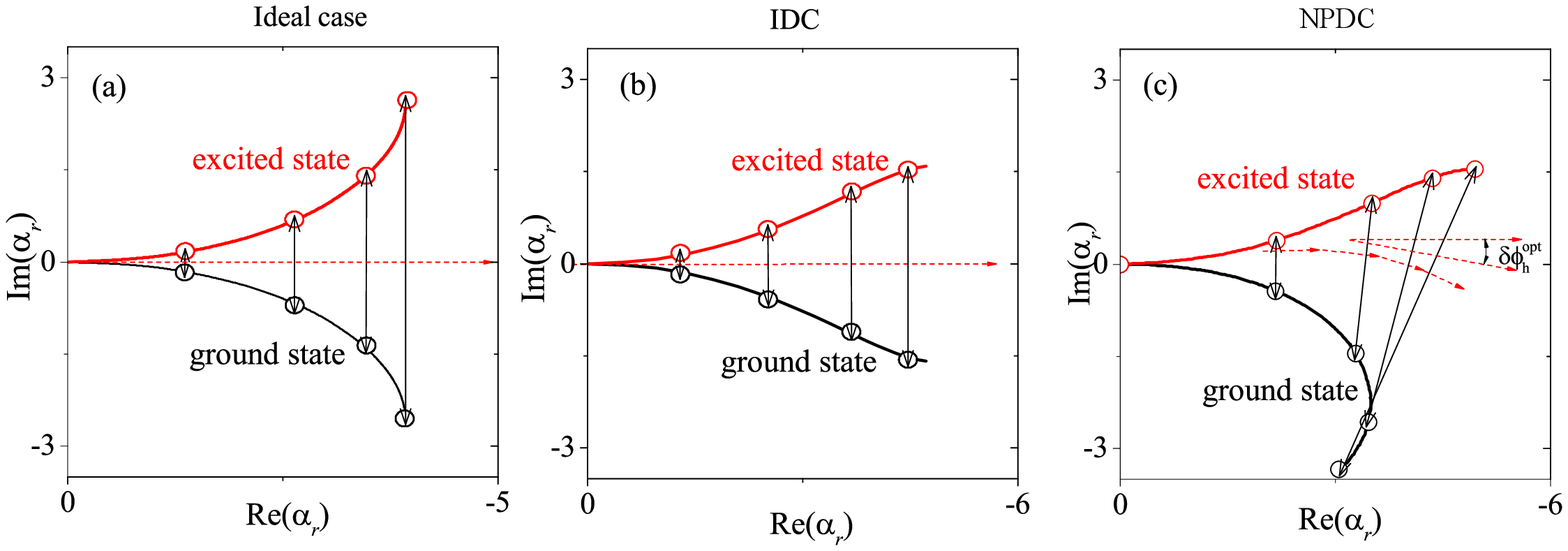}
    \caption{Evolutions in the phase space of the intracavity field
        for (a) the ideal [(Eq.~\ref{alphabar})],
        (b) IDC [(Eq.~\ref{inducedLan})],
        and (c) NPDC [(Eq.~\ref{directLan})] readouts.
        The red (black) curves correspond to the qubit being in its excited (ground) state,
        and the black solid arrows connecting two circles represent the separations
        between two signals at the same time
        $t$. In (c), $\delta\phi_{h}^{\text{opt}}$ indicates the rotated
        optimal homodyne angle for a homodyne measurement.
        In these plots, both Kerr nonlinearity and Purcell effects are considered.
        The parameters used here are the same as those in Fig.~\ref{fig4m}(d),
        and the drive strength is assumed to be the same as for the stop point.
    }
    \label{fig4s}
\end{figure*}

Due to the dispersive coupling, the qubit in its ground or excited
states corresponds to rotating the output field in phase space
with two different angles. The qubit state is encoded in the
output quadrature
$Y(\phi_{h})=a_{\text{out}}^{\dag}e^{i\phi_{h}}+a_{\text{out}}
e^{-i\phi_{h}}$ with $\phi_{h}$ being the homodyne-measurement
angle. The output signal corresponds to a standard homodyne
detection of the quadrature $Y(\phi_{h})$, with an integration time
$\tau$, and has the following form
\begin{equation}
M(\tau) =\sqrt{\kappa} \int_{0}^{\tau} dt
\left[a_{\text{out}}^{\dag}(t)\exp(i\phi_{h})+a_{\text{out}}(t)
\exp(-i\phi_{h})\right], \label{signal}
\end{equation}
By setting $\langle \sigma_{z}\rangle =\pm1$ in
Eq.~(\ref{alphabar}), respectively, one obtains the expression for
the separation signal given in Eq.~(\ref{signalS}).
On the other hand, the fluctuations $d_{\text{out}}(t)$ brings
noise into the measurement signal. The integrated imprecision
noise $M_{N}(\tau)$ is identical for the qubit ground and excited
states, and is expressed as~\cite{Didier152}
\begin{eqnarray}
M_{N}^{2}(\tau)&=&[\langle M_{N}^{2}(\tau)\rangle
_{|e\rangle}+\langle M_{N}^{2}(\tau)\rangle _{|g\rangle}] \notag   \\
&=&2\kappa\left\{\int_{0}^{\tau}\!\! dt
[d_{\text{out}}^{\dag}(t)e^{i\phi_{h}}\!+\text{H.c.}]\right\}^{2}
\!=\!2\kappa \tau. \label{noise}
\end{eqnarray}

According to Eq.~(\ref{signalS}), the signal $M_{s}(\tau)$ is
optimized by setting $\phi_{h}'=\theta_{d}-\phi_{h}=\pi/2$ and
$\theta_{q}=\pi/4$ (i.e., $\chi_{z}=\kappa/2$) in the long-time
limit with $\kappa \tau\gg 1$. In Fig.~\ref{fig4s}(a),  by
adopting the same parameters as those in Fig.~3(d) in the main
article (the drive strength is assumed at the stop point), we plot
the evolution of the intracavity fields in phase space. The
red and black curves represent the qubit in its excited and ground
states, respectively. The two circles connected by the same black
arrow correspond to the same time $t$. We find that the
separation direction between these two signals in phase space
is along the black solid arrows and is always vertical to the red
dashed arrow, which corresponds to the optimal relative angle
$\phi_{h}'=\theta_{d}-\phi_{h}=\pi/2$ of a homodyne measurement.
The signal-to-noise-ratio (SNR) becomes
\begin{eqnarray}
\mathcal{R}&=&\frac{M_{s}(\tau)}{M_{N}(\tau)} \notag \\
&=&\frac{2\epsilon\sqrt{2\kappa\tau}}{\kappa}\!\left[\!1\!-\!\frac{2}{\kappa\tau}\left(1-e^{-\frac{1}{2}\kappa
    \tau}\cos\frac{1}{2}\kappa\tau \right)\right]. \label{SNR}
\end{eqnarray}
In the following discussion, we discuss the IDC and NPDC readouts.
We find that the optimal measurement signal described in
Eq.~(\ref{SNR}) can be destroyed by both the Kerr and Purcell
effects.

\subsection*{E.2 Kerr nonlinearity for the IDC- and NPDC-based readout mechanisms}

\begin{figure*}[tbp]
    \centering
    \includegraphics[width=17.7cm]{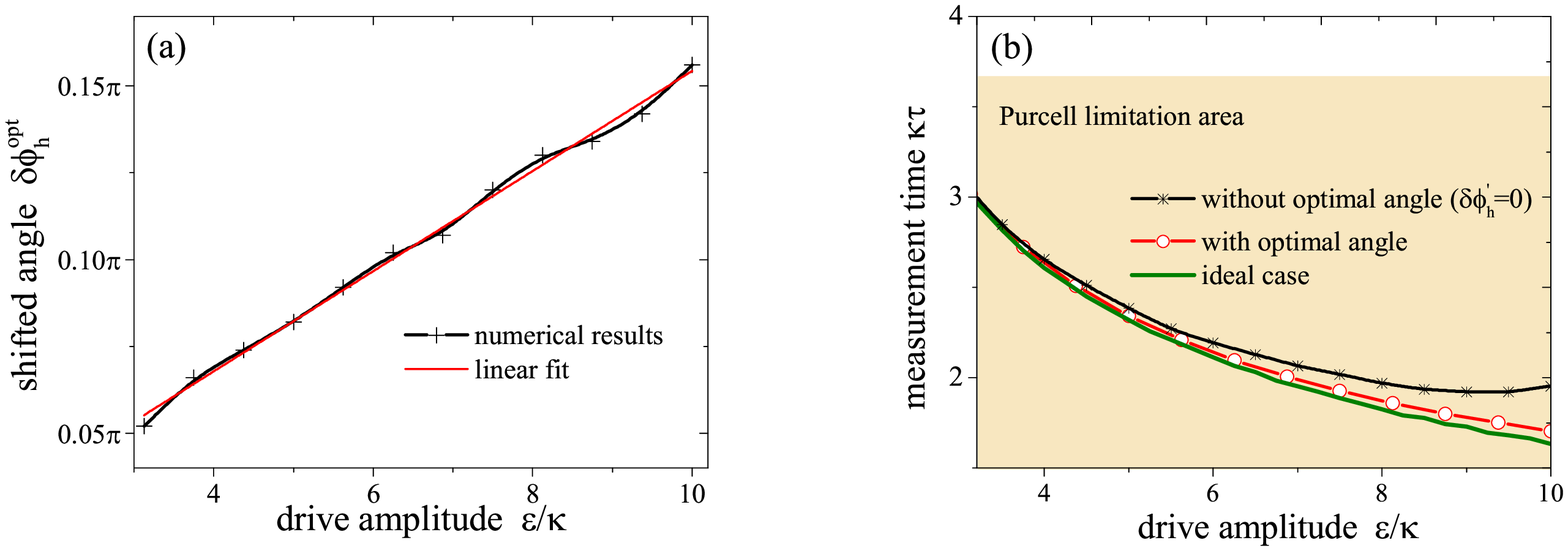}
    \caption{
        (a) Optimal shifted homodyne angle $\delta\phi_{h}^{\text{opt}}$
        (corresponding to the shortest measurement time to reach $F=99.99\%$)
        versus the drive strength $\epsilon$ (in units of $\kappa$).
        The curve with crosses shows the numerical results, and the red
        curves show our linear fit according to the relation
        $\epsilon/\kappa=a \delta\phi_{h}^{\text{opt}}+b$. Here we set
        $a\simeq0.144$ and $b\simeq0.01$. (b) Required measurement time to
        reach the fidelity $F=99.99\%$ for the ideal readout
        [Eq.~(\ref{SNR})], the NPDC readouts with and without the shifted
        optimal angle $\delta\phi_{h}^{\text{opt}}$. The cyan area
        corresponds to the Purcell limitation area. The parameters here are the same as those in Fig.~\ref{fig4m}.}
    \label{fig5s}
\end{figure*}

In the IDC readout, according to the nonlinear Langevin
equation~(\ref{inducedLan}), the effective cavity frequency pull
$\xi_{g}(t)$ is reduced by the photon number due to the
qubit-dependent Kerr terms, which can be written as:
\begin{eqnarray}
\xi_{g}(t)&=&-\chi_{z}-2Kn_{g}(t)=-\chi_{z}\left[1- \frac{\langle
    n_{g}(t) \rangle}{2n_{c} } \right],   \\
\xi_{e}(t)&=&\chi_{z}+2Kn_{e}(t)=\chi_{z}\left[1- \frac{\langle
    n_{e}(t) \rangle}{2n_{c} } \right], \label{cavitypull}
\end{eqnarray}
where $\langle n_{g,e}(t)\rangle=\langle
a^{\dagger}(t)a(t)\rangle\Big|_{g,e}$ is the mean photon number
when the qubit is in its ground and excited states, respectively.
As discussed in Ref.~\cite{Boissonneault09},
Eq.~(\ref{cavitypull}) indicates that with increasing the
measuring photon number, the effective cavity pull $\xi$ is
decreased. Specifically, when the intracavity number reaches
$\langle n_{g,e}(t) \rangle=n_{c}$, the cavity pull is reduced as
$\xi_{g,e}(t)=\chi_{z}/2$.

Because the qubit-dependent Kerr nonlinearity is symmetric for the
ground and excited states, it can be easily verified that $\langle
n_{g}(t) \rangle\simeq \langle n_{e}(t) \rangle$ and
$\xi_{g}(t)\simeq -\xi_{e}(t)$.
The reduction of the cavity pull $\xi_{g,e}(t)$ reduces the signal
separation in phase space, which can be clearly found by
comparing the numerical results in Figs.~\ref{fig4s}(b) and
\ref{fig4s}(a). With increasing time $t$, the separation distances
(the black arrows) are significantly reduced compared with those
in the ideal readout. Consequently, the required measurement time
becomes longer. Therefore, for the IDC readout, increasing the
intracavity photon number does not only enhance the
qubit-error-transition probability (Purcell photon number
limitations), but also reduces the measurement fidelity due to the
Kerr nonlinearity $K$.

For the NPDC readout without the dipole-field coupling, because
the intracavity photons do not deteriorate the qubit states, there
is no qubit-error-transition due to the Purcell effects. However,
when $\langle n(t)\rangle$ is large, the Kerr nonlinearity
(introduced by the SQUID) induces apparent effects. Different from
the IDC readout, the nonlinearity is not qubit-dependent, and the
changing of the cavity pull for the two qubit states is not
symmetric. For $K<0$, the cavity pulls for the qubit being in its
ground and excited state are, respectively,
\begin{equation}
\xi_{g}(t)=-\chi_{z}+2Kn_{g}(t), \quad
\xi_{e}(t)=\chi_{z}+2Kn_{e}(t),
\end{equation}
from which we can find that, by increasing the photon number, the
effective cavity pull $|\xi_{g}|$ ($|\xi_{e}|$) increases
(decreases). This leads to asymmetric rotation angles of the
cavity field for the qubit in the ground and excited states, which
can be clearly found from Fig.~\ref{fig4s}(c). The evolutions for
the ground and excited states in phase space are asymmetric. The
signal separation direction (black arrows) is now time-dependent
and rotates in phase space. However, the signal separation
distance is still large compared with the NPDC readout, and almost
equal to that in the ideal readout.

In a homodyne experiment, one can tune the measurement angle
$\phi_{h}'$ for the NPDC readout, to maximize the total signal
separation $M_{s}(\tau)$ during the integrating time $\tau$. As
sketched in Fig.~\ref{fig4s}(c), $\phi_{h}'$ can only be slightly
shifted with an amount $\delta\phi_{h}'$. For a certain drive
strength $\epsilon$, there exists an optimal shifted angle
$\delta\phi_{h}^{\text{opt}}$, which corresponds to the shortest
measurement time for a certain fidelity. In Fig.~\ref{fig5s}(a),
by adopting the same parameters as those in Fig.~3 of the main
article (with $F=99.99\%$), we plot $\delta\phi_{h}^{\text{opt}}$
changing with $\epsilon$. It can be found that, with a stronger
drive strength $\epsilon$, we need a larger optimal shifted angle
$\delta\phi_{h}^{\text{opt}}$. Their relation can be approximately
described by a simple linear function $\epsilon/\kappa=a
\delta\phi_{h}^{\text{opt}}+b$ (red curve). In
Fig.~\ref{fig5s}(b), we plot the required measurement time to
reach the fidelity $F=99.99\%$ for the ideal readout, the NPDC
readout with and without shifted optimal angle
$\delta\phi_{h}^{\text{opt}}$. We find that, even without shifted
optimal angle (curve with asterisks), the measurement can still go
into the Purcell limitation area of the IDC readout (yellow area,
below the stop point in Fig.~\ref{fig4m}). If we choose
the optimal shifted angle $\delta\phi_{h}^{\text{opt}}$, the
required time can still be shortened (curve with circles), and it
is close to that of the ideal readout (green solid curve). Therefore, by
slightly rotating the measurement angle, the measurement time can
go far below the Purcell limitation area compared with the IDC
readout.

\end{appendix}
%

\end{document}